# Escape from crossover interference increases with maternal age

Authors: Christopher L. Campbell[1], Nicholas A. Furlotte[2], Nick Eriksson[3], David Hinds[2], Adam Auton[1, *]

1. Albert Einstein College of Medicine, Department of Genetics, 1301 Morris Park Ave., Bronx, NY 10461, USA
2. 23andMe, Mountain View, California, USA
3. Former employee of 23andMe

* Corresponding Author. Email: adam.auton@einstein.yu.edu

## Abstract

Recombination plays a fundamental role in meiosis, ensuring the proper segregation of chromosomes and contributing to genetic diversity by generating novel combinations of alleles. Using data derived from direct-to-consumer genetic testing, we investigated patterns of recombination in over 4,200 families. Our analysis revealed a number of sex differences in the distribution of recombination. We find the fraction of male events occurring within hotspots to be 4.6% higher than for females. We confirm that the recombination rate increases with maternal age, while hotspot usage decreases, with no such effects observed in males. Finally, we show that the placement of female recombination events becomes increasingly deregulated with maternal age, with an increasing fraction of events appearing to escape crossover interference.

## Main Text

Recombination is a fundamental meiotic process that is required to ensure the proper segregation of chromosomes. In mammals and other eukaryotes, at least one crossover is normally required to ensure proper disjunction, and failures in recombination can result in deleterious outcomes such as aneuploidy. As such, the recombination process is highly regulated in order to ensure that sufficient numbers of crossovers occur. The placement of crossover events along a chromosome is also tightly regulated. At the fine scale, the majority of crossovers tend to occur within localized regions of approximately 2kb in width known as recombination hotspots. At broader scales, interference between crossovers appears to increase spacing between events occurring on the same chromosome during meiosis.

As relatively few crossover events occur within a single meiosis, quantifying the recombination landscape requires the observation of large numbers of meioses. To investigate properties of crossover placement in humans, we collected data from pedigree families contained within the database of 23andMe, Inc. (Mountain View, CA). Our dataset consists of 4,209 families contributing a total of 18,302 informative meioses genotyped at over 515,972 sites. To preserve the privacy of the participants, families were removed if the age of the mother was greater than 40 years at the time of childbirth, the age of the father was greater than 45 years, or the difference between the parental ages was greater than 15 years (Supplementary Figure 1). The majority of the data is derived from family quartets (Supplementary Table 1), accounting for 78.6% of the families, and is also predominately composed of individuals of European ancestry (Supplementary Table 2). Ancestral populations are assigned to each individual by comparison to a set of reference populations (see Supplementary Material).

To infer recombination events in nuclear families, we applied the Lander-Green algorithm as implemented within Merlin (version 1.1.2; *1*). To guard against genotyping error, we curated the data to remove nearby recombination events that could be indicative of genotyping error (see Supplementary Material; Supplementary Figure 2). This approach allowed us to identify over 645,000 well-supported crossover events, with the median event being localized to 28.2 kb (Supplementary Figure 3).

We inferred a mean of 41.6 autosomal recombination events per gamete in females (95% confidence interval (CI): 41.4-41.9), and 26.6 in males (95% CI: 26.5-26.7, Figure 1A). The genetic map constructed from our data agrees well with those generated by previous studies (Figure 1B, Supplementary Figure 4, Supplementary Table 3). At the 5Mb scale, the Pearson correlation between our map and that of deCODE (*2*) is $r^2 = 0.975$ and 0.983 for females and males respectively. Likewise, our sex-averaged map has a correlation of $r^2 = 0.955$ with the HapMap map inferred from patterns of linkage disequilibrium (LD) (*3*). At the chromosome scale, the map length is well predicated by the physical chromosome length ($r^2 = 0.991$ in females and 0.945 in males; Supplementary Figure 5).

Treating the overall recombination rate as a phenotype, we replicate genetic associations at genome-wide significance for RNF212, which is known to be essential for crossover-specific complexes (*4*), and within the vicinity of TTC5, which appears to replicate an association with CCNB1IP1 (*5*). Another association near SMEK1 also replicates discoveries elsewhere (*5*), but not at genome-wide significance (Supplementary Table 4).

Previous reports have suggested increased recombination rates in older females (*6, 7*). Using linear regression (Supplementary Figure 6), we obtain a similar result with an additional 0.067 events per year being observed in females ($p = 0.002$), and no such effect being observed in males ($p = 0.30$). The female effect appears to be driven by sharp increase in the number of recombination events for older mothers (Figure 1C). Fitting piecewise-linear model with a single change-point infers a rapid increase in the female recombination rate after 38.8 years, increasing from 0.047 events per year to 2.990 events per year. On average, mothers of 39 years and over have an additional 2.51 events compared to younger mothers ($p = 0.0005$, Mann-Whitney U).

Both pedigree and LD studies have suggested that approximately 60-70% of crossover events occur within recombination hotspots (*7, 8*). Our data confirms this result, with 62.7% of events occurring within LD-defined hotspots in females, and

67.3% occurring within hotspots in males (Figure 2A; Supplementary Figure 7A). The 4.6% difference between the two sexes is highly significant ($p = 1.1\times10^{-69}$), suggesting differences in the regulation of crossover placement between the sexes. The result remains significant after thinning the female data to match the crossover density of the male data ($p < 2.2\times10^{-16}$; see Supplementary Material), and does not appear to be driven by increased male recombination rates near the telomeres (see Supplementary Material).

Hotspot localization is believed to be under the control of the zinc-finger protein PRDM9, which recognizes and binds specific DNA motifs (*9-11*). We find SNPs in the vicinity of PRDM9 to be strongly associated with the degree of hotspot usage, as has previously been reported (*5, 11*). The most strongly associated SNP is rs73742307 achieving a p-value of $7.9 \times 10^{-184}$, with no other region achieving a genome-wide significant association with this phenotype (Supplementary Table 5).

Variation within the PRDM9 DNA-binding domain can result in changes to the recognized motif and hence lead to differences in hotspot localization between individuals. While the major allele of PRDM9 (allele A) is present at high frequency in most human populations, a large number of low frequency alleles have been observed, particularly within African populations (*10, 12*). Consistent with this, we find hotspot usage to be significantly lower within individuals of African ancestry (Figure 2B, Supplementary Table 6), which reflects the fact that the LD-defined hotspots are expected to mostly represent the common PRDM9 allele. Notably, while over 75% of our data is derived from individuals of European ancestry, hotspot usage is higher for males than females across all ancestries.

We find a weak association between hotspot usage and maternal age (Supplementary Figure 7B). Using logistic regression, we estimate a decrease in hotspot usage corresponding to ~1% over a 10 year period ($\beta_1 = -0.0042$, s.e. $= 9.6\times10^{-4}, p = 1.2\times10^{-5}$). To ensure this effect is not driven by differences in parental ancestry within the sample, we repeated the analysis only using individuals of European ancestry. In this case, the effect size remains similar ($\beta_1 = -0.0033$, s.e. $= 0.0013$), but is only marginally significant ($p = 0.0101$). Including the number of events as an additional predictor variable within the regression leaves age as a weakly significant predictor ($p = 0.0106$), but not the number of events ($p = 0.74$). Despite the small size of the estimated effect, we note that no such age-related effects were observed in males.

To learn more about interactions between recombination events, we used the high number of crossover locations in our data to better characterize the phenomenon of crossover interference. By considering the distribution inter-crossover distances, we fit three models to describe the distribution of inter-crossover distances: a model without interference between crossovers, the gamma model of crossover interference (*13*), and a gamma mixture model in which a subset of events escape interference (*14*). We refer to these three models respectively as the 'interference free' model, the 'simple interference' model, and the 'interference escape' model.

In agreement with previous reports (*14, 15*), the interference escape model provides a much better fit to our data than either the simple interference or interference free models (Figure 3A). Under this model, the estimates of the strength of crossover interference are similar to previous reported using smaller data sets (*15*). The degree of interference is inferred to be lower in females than in males ($\nu_{female} = 7.19$ vs $\nu_{male} = 8.93$). In addition, 7.8% / 6.7% of female / male events are inferred to escape interference. We therefore conclude that a non-negligible fraction of crossovers occur in the absence of crossover interference.

We find evidence that both the degree of interference and interference escape varies across chromosomes (Figure 3B and C, Supplementary Table 7). The strength of interference is reasonably well predicted by the chromosome map length ($r^2 = 0.565, p = 6.4 \times 10^{-9}$), although the relationship is only significant in females when considering the sexes separately ($r^2_{female} = 0.69, p = 1.7 \times 10^{-6}$ and $r^2_{male} = 0.172, p = 0.06$; Supplementary Figure 8). In contrast, the fraction of events escaping interference shows no relationship with chromosome map length ($r^2 = 0.001, p = 0.84$). Certain chromosomes appear to have high degrees of escape, with chromosomes 8, 9 and 16 (in females) being notable outliers.

To investigate if crossover interference changes with parental age, we subdivided our data into 10 quantiles on the basis of age, and fit the interference escape model for each group independently. We observe a striking increase in the proportion of events that escape interference with maternal age (Figure 4A), rising from 6.7% for mothers under 25 years to 9.5% for mothers over 35 years. No such correlation is observed for the interference parameter in females, and no correlation is observed for either parameter in males (Supplementary Figure 9). The effect is robust different subdivisions of the data (Supplementary Figure 10 and Supplementary Figure 11).

A potential concern is that the detected increase in interference escape could be driven by the observed increased number of crossovers in older mothers. If the number of crossovers is increased, then the distances between them are necessarily shorter, which may in turn influence the interference parameter estimates. To account for this possibility, we performed stratified sampling of individuals to control for the number of events within each quantile. The observed increase in of the escape parameter with maternal age is still observed (Supplementary Figure 12), indicating that it is not driven by changes in the overall recombination rate.

To further investigate the differences between old and young parents, we plotted the distribution of inter-crossover distances for young and old parents (Figure 4B and C). The interference-escape effect in females appears to be predominately driven by an increase in the number of very tightly clustered events, generally separated by less than ~5 cM. These tightly clustered events are not well modeled by the interference escape model (Supplementary Figure 13), and a major concern therefore is that these tightly clustered events represent false positive calls arising from genotyping error. However, the effect remains even if we apply much stricter filtering of the crossover events (Supplementary Figure 14), and in addition we believe genotyping error is unlikely to explain the association between the escape parameter and maternal age because a) the effect is not seen in males, and b) it would imply increased genotyping error for older mothers (but not fathers).

In terms of meiosis, a major difference between the sexes is that female meiosis starts during fetal development, but does not complete until adulthood. As such, while male gametes are produced throughout adulthood and promptly proceed through meiosis, oocytes remain arrested in a late stage of prophase (dictyotene) for many years, if not decades. Presuming our observation of increasing crossover interference escape with maternal age is not due to some obscure form of genotyping error, our observations add to similar evidence of increasing rates of recombination (*6*) and aneuploidy (*16*) in aging females. Although these phenomena are presumably related, the biological mechanisms by which they occur are unclear, and we can think of at least three possibilities. First, given chromatids remain physically proximal during the extended period of female meiotic arrest, one possible explanation is that additional recombinations are initiated during this time, perhaps in response to DNA damage. However, as recombination is believed to have completed by the time of dictyotene, such an explanation appears unlikely. A second possibility, previously invoked to explain the increasing recombination rate with maternal age (*6*), suggests oocytes with additional recombination events could be at reduced risk of nondisjunction, and hence would be more likely to lead to viable embryos in older mothers. However, it is not clear that this mechanism would explain the increased clustering of events observed in our data. Finally, a third possibility is related to the so-called "production-line" hypothesis, in which oocytes are selected for maturation sequentially in the same order as their generation, and later oocytes have therefore potentially undergone additional mitotic divisions prior to entering meiosis (*17*). However, the existence of a production line has been debated for many years (*17-19*), and so the likelihood of this explanation is unclear.

## Author contributions

Designed study: A.A., D.H, N.E. Conducted analysis: A.A., C.C, N.F. Wrote paper: A.A, C.C.

## Figure Legends

Figure 1: A) The number of events per meiosis for females (red) and males (blue), with median values indicated by a vertical line. For phase-unknown individuals, the average number of events per meiosis was used. B) Squared Pearson correlation between the 23andMe map, the deCODE map, and the HapMap map, as a function of scale. C) The number of recombination events as a function of parental age for females (red) and males (blue), relative to parents of between 20 and 25 years of age. Parents were grouped into 5-year age bins, and the mean number of events estimated. Error bars show a 95% confidence interval for each group.

Figure 2: A) Hotspot usage for female (red) and male (blue) meioses. Median values for each sex are shown by vertical lines. B) Mean hotspot usage, subdivided by parental population. Females are shown in red, males in blue, and a combined estimate in black. Error bars indicate a 95% confidence interval.

Figure 3: A) Fit of three models of interference to the inter-crossover distances observed on chromosome 1, derived from phase-known mothers (red) and fathers (blue). The interference free model is shown as a dotted line, the simple interference model is shown as a dashed line, and the interference escape model is shown as a solid line. B) Per-chromosome estimates of the interference parameter as estimated from the interference escape model. Error bars indicate a 95% confidence interval. Note that chr21 in males is excluded due to an extremely high estimate. C) Per-chromosome estimates of the proportion of events escaping interference. Error bars indicate a 95% confidence interval.

Figure 4: A) Inferred escape parameter as a function of maternal age. Mothers were divided into 10 approximately equal sized deciles on the basis of age, and the interference escape model was fitted for each group separately. Estimates for $\nu$ show no correlation with age (Supplementary Figure 9). Error bars indicate 95% confidence intervals. B) Distribution of inter-crossover distances for young and old mothers, where the boundary between young and old is taken as median maternal age (30 years). Error bars represent a 95% confidence interval assessed via 1000 bootstrap samples, and the arrow highlights a significant difference between the young and old groups for tightly clustered events. The insert shows that cumulative distribution function up to 5 cM. C) Distribution of inter-crossover distances for young and old fathers, where the boundary between young and old is taken as median paternal age (32 years).

## Supplementary Figure Legends

Supplementary Figure 1: Age distributions within the filtered dataset. The left hand panel shows the distribution for phase unknown individuals, where the parental ages were averaged across children. The right hand panel shows data for the phase known meioses where the parental age at the time of childbirth is known. Lines indicate the mean of each distribution. Note that some families were excluded from analysis by 23andMe on the basis age to protect privacy, as seen from the truncated distribution of maternal ages in the right hand panel.

Supplementary Figure 2: Data grooming. A) Chromosome 10 map before filtering. Genetic maps from the 23andMe data are shown in bold lines, whereas the genetic maps from deCODE are shown as thin lines. Separate maps are shown for females (red), males (blue), and sex-averaged (black). Also shown are regions highlighted in grey that represent gaps in the reference assembly, the largest of which being the centromere at around 40Mb. B) Clustering of recombination events occurring within 1Mb of each other within single individuals. Each plot shows the number of events within 1Mb of each other on a $\log_{10}$ scale as a function of physical position on each chromosome. A large number of these event pairs can be observed on chromosome 10, although other large peaks can also be observed on, for example, chromosomes 8 and 15. The dashed line represents the 99.9% percentile of the distribution, and was used as a threshold for filtering. C) Chromosome 10 map after filtering.

Supplementary Figure 3: Empirical cumulative distance function of crossover localization distances. Red labels indicate the interval distances at the distribution deciles.

Supplementary Figure 4: Genetic map estimated from 23andMe data. Genetic maps from the 23andMe data are shown in bold lines, whereas the genetic maps from deCODE are shown as thin lines. Separate maps are shown for females (red), males (blue), and sex-averaged (black). Also shown are regions highlighted in grey that represent gaps in the reference assembly. For PAR1, we are showing data derived from Duffy (*20*) for comparison. As the deCODE maps cover a slightly smaller physical region than the 23andMe maps, the deCODE maps have been shifted slightly upwards to aid visual comparison. Specifically, the deCODE map has been aligned with the 23andMe map at the first physical position within the deCODE map. The locations of the alignments are indicated by small circles that can be most clearly seen on the smaller chromosomes.

Supplementary Figure 5: The relationship between chromosome length and recombination. The top row shows the correlation between physical length and map length for females (left), males (center), and sex averaged (right), with a linear fit included for the 23andMe map (red) and the deCODE map (blue). The bottom row shows the relationship between physical length and average recombination rate with a quadratic fit. Note that chromosome X has been included in the female plots, but was excluded from the regressions.

Supplementary Figure 6: Number of autosome recombination events verses parental age for females (left) and males (right). A linear least-squares fit is indicated by a black line. The least-squares fit equation given in the legend together with a p-value for the non-constant term.

Supplementary Figure 7: A) Hotspot usage estimated in females (left) and males (right). The MLE estimate for each individual is indicated by a circle, with a 95% confidence interval indicated by the shaded area. The median MLE estimate for each sex is indicated by a vertical black line. B) Hotspot usage by parental age for females (left) and males (right). For each plot a logistic regression is also shown, with the p-value for the non-constant term given in the title.

Supplementary Figure 8: A) The relationship between chromosome map length and the interference parameter, *v*. B) The relationship between chromosome map length

and the escape parameter, p. Linear fits are shown for females (red), males (blue), and the data combined across sexes (black). In both plots, the chr21 estimate in males has been excluded.

Supplementary Figure 9: Interference parameters as a function of age. Females and males are shown on the top and bottom rows respectively. Estimates of the interference parameter, $v$, are shown on the left, whereas estimates of the escape parameter, $p$, are shown on the right. Error bars show 95% confidence intervals.

Supplementary Figure 10: Interference parameters by age, having divided the data in 5 or 20 age quantiles. Error bars show 95% confidence intervals.

Supplementary Figure 11: Interference parameters by age, having estimated the interference parameters for phase-known and phase-unknown groups separately.

Supplementary Figure 12: Interference parameters as a function of age, following stratified sampling. Females and males are shown on the top and bottom rows respectively. Estimates of the interference parameter, $v$, are shown on the left, whereas estimates of the escape parameter, $p$, are shown on the right. Error bars show 95% confidence intervals.

Supplementary Figure 13: Model fit for tightly clustered events in females (A) and males (B). The figure shows the empirical cumulative distribution function for young (green line) and old (magenta line) mothers/fathers, and compares to that obtained via simulation under the interference free model (black dotted line), the simple interference model (black dashed line), and the interference escape model (solid black line), with parameters were taken from Supplementary Table 7. The figure is shown on a log-log scale to emphasize the short inter-crossover distances.

Supplementary Figure 14: Interference parameters estimated for a strictly filtered dataset. In this case, all crossover events were required at least 10 supporting informative sites (compared to 3 in the main dataset), no two events within a single family were allowed to be within 5 SNPs of each other (compared to 1 in the main dataset), and no more than 4 events within 1Mb of each other were allowed across the whole dataset (and compared to 14 in the main dataset, which corresponds to the 99.9[th] percentile). After this very strict filtering, the deviation from the interference escape model is much less pronounced at short scales (right hand panels), but the association between interference escape and maternal age remains strong (2[nd] panel from top left).

## Supplementary Tables

| Pedigree Type | Description | Before Filtering | After Filtering |
|---|---|---|---|
| 1 | 2 parents, 2 children | 3319 | 3307 |
| 2 | 2 parents, 3 children | 560 | 523 |
| 3 | 2 parents, 4 children | 89 | 80 |
| 4 | Quartet, with 2nd generation trio | 101 | 100 |
| 5 | Trio, with 2nd generation quartet | 201 | 199 |
| | Total | 4270 | 4209 |

Supplementary Table 1: Summary of dataset, before and after filtering.

| Population | Female unphased | Male unphased | Female phased | Male phased | Total Meioses | Percentage |
|---|---|---|---|---|---|---|
| Europe | 5382 | 5508 | 1789 | 1641 | 14320 | 78.24% |
| Latino | 602 | 546 | 171 | 190 | 1509 | 8.25% |
| East Asia | 380 | 308 | 88 | 74 | 850 | 4.64% |
| None/Other | 198 | 268 | 68 | 109 | 643 | 3.51% |
| South Asia | 178 | 176 | 19 | 20 | 393 | 2.15% |
| African American | 152 | 152 | 34 | 36 | 374 | 2.04% |
| Middle East | 76 | 100 | 15 | 22 | 213 | 1.16% |
| Total | 6968 | 7058 | 2184 | 2092 | 18302 | 100.00% |

Supplementary Table 2: Description of parental ancestry for each meiosis within the sample.

| Chrom | First Position (bp) | Last Position (bp) | Physical Length (Mb) | Female Map Length (cM) | Female Mean Rate (cM/Mb) | Male Map Length (cM) | Male Mean Rate (cM/Mb) | SexAvg Map Length (cM) | SexAvg Mean Rate (cM/Mb) |
|---|---|---|---|---|---|---|---|---|---|
| chr1 | 1,031,540 | 249,170,711 | 248.14 | **335.90** | 1.36 | **198.30** | 0.80 | **267.05** | 1.08 |
| chr2 | 118,913 | 242,763,542 | 242.64 | **316.45** | 1.31 | **184.64** | 0.76 | **250.52** | 1.03 |
| chr3 | 152,592 | 197,759,785 | 197.61 | **270.98** | 1.37 | **163.85** | 0.83 | **217.40** | 1.10 |
| chr4 | 167,596 | 190,787,660 | 190.62 | **260.11** | 1.37 | **145.79** | 0.76 | **202.93** | 1.06 |
| chr5 | 184,702 | 180,673,228 | 180.49 | **249.13** | 1.38 | **146.66** | 0.81 | **197.87** | 1.10 |
| chr6 | 188,937 | 170,777,087 | 170.59 | **236.64** | 1.39 | **140.88** | 0.83 | **188.74** | 1.11 |
| chr7 | 67,365 | 159,042,351 | 158.97 | **223.17** | 1.41 | **136.04** | 0.86 | **179.55** | 1.13 |
| chr8 | 200,898 | 146,235,564 | 146.03 | **210.94** | 1.45 | **122.41** | 0.84 | **166.64** | 1.14 |
| chr9 | 215,269 | 141,004,945 | 140.79 | **195.69** | 1.40 | **125.54** | 0.89 | **160.58** | 1.14 |
| chr10 | 162,102 | 135,402,200 | 135.24 | **207.86** | 1.54 | **129.91** | 0.96 | **168.86** | 1.25 |
| chr11 | 244,552 | 134,872,342 | 134.63 | **193.59** | 1.44 | **120.21** | 0.89 | **156.88** | 1.17 |
| chr12 | 216,039 | 133,684,321 | 133.47 | **200.36** | 1.51 | **131.20** | 0.98 | **165.75** | 1.24 |
| chr13 | 19,458,371 | 114,998,076 | 95.54 | **152.26** | 1.60 | **101.19** | 1.06 | **126.71** | 1.33 |
| chr14 | 20,445,905 | 107,233,999 | 86.79 | **137.22** | 1.59 | **97.29** | 1.12 | **117.24** | 1.35 |
| chr15 | 22,763,396 | 102,381,360 | 79.62 | **143.39** | 1.80 | **100.85** | 1.27 | **122.11** | 1.53 |
| chr16 | 143,503 | 90,102,384 | 89.96 | **157.29** | 1.75 | **102.03** | 1.13 | **129.64** | 1.44 |
| chr17 | 84,782 | 81,025,393 | 80.94 | **152.87** | 1.90 | **106.23** | 1.31 | **129.53** | 1.60 |
| chr18 | 218,695 | 77,955,378 | 77.74 | **140.06** | 1.81 | **97.80** | 1.26 | **118.91** | 1.53 |
| chr19 | 288,246 | 59,058,083 | 58.77 | **117.80** | 2.01 | **99.42** | 1.69 | **108.59** | 1.85 |
| chr20 | 100,699 | 62,892,739 | 62.79 | **118.90** | 1.90 | **99.00** | 1.58 | **108.93** | 1.73 |
| chr21 | 14,807,136 | 47,978,421 | 33.17 | **74.34** | 2.24 | **51.76** | 1.58 | **63.04** | 1.90 |
| chr22 | 17,152,611 | 51,165,664 | 34.01 | **78.16** | 2.31 | **63.30** | 1.86 | **70.71** | 2.08 |
| chrX | 2,737,282 | 154,408,041 | 151.67 | **179.02** | 1.18 | | | | |
| PAR1 | 178,624 | 2,689,575 | 2.51 | **2.73** | 1.16 | **42.94** | 17.17 | **22.75** | 9.06 |
| PAR2 | 154,984,651 | 155,227,607 | 0.24 | **0.05** | 0.34 | **0.33** | 1.35 | **0.19** | 0.79 |
| **Genome** | | | 2932.98 | 4354.91 | 1.48 | 2707.55 | 0.92 | 3441.11 | 1.17 |

Supplementary Table 3: Properties of the map estimated from 23andMe data. Recombination fractions were converted to genetic map distances using the Haldane map function.

| SNP | Chrom | Position | Alleles | P-value | Effect | 95% CI | Gene Context |
|---|---|---|---|---|---|---|---|
| rs2001572 | chr14 | 20,767,868 | A/T | 1.50E-08 | 0.503 | [0.329,0.677] | [TTC5] |
| rs79621814 | chr4 | 1,089,268 | C/T | 2.90E-08 | -0.99 | [-1.340,-0.640] | [RNF212] |
| rs11624006 | chr14 | 91,961,188 | C/T | 2.80E-07 | -0.478 | [-0.660,-0.296] | [SMEK1] |
| rs72631326 | chr17 | 65,769,087 | C/T | 4.40E-07 | 0.959 | [0.587,1.331] | NOL11--[]--BPTF |
| rs11932663 | chr4 | 184,458,083 | A/G | 5.10E-07 | 0.622 | [0.380,0.865] | ING2--[]---RWDD4 |
| rs17127442 | chr8 | 18,779,787 | C/T | 5.10E-07 | -0.537 | [-0.746,-0.327] | [PSD3] |
| rs1879904 | chr11 | 82,076,387 | C/T | 6.80E-07 | -0.507 | [-0.707,-0.307] | []---FAM181B |

Supplementary Table 4: Variants associated with total number of recombination events. Linear regression model tested as N_events ~ sex + age + pc.0 + pc.1 + pc.2 + pc.3 + pc.4 + genotype. Association tests conducted using only individuals found to have ≥ 97% European ancestry.

| SNP | Chrom | Position | Alleles | P-value | Effect | 95% CI | Gene Context |
|---|---|---|---|---|---|---|---|
| rs73742307 | chr5 | 23,534,421 | C/T | 7.90E-184 | 0.16 | [0.149,0.170] | PRDM9-[]---CDH10 |
| rs78474856 | chr20 | 1,450,623 | C/G | 6.10E-07 | -0.021 | [-0.029,-0.013] | NSFL1C-[]-SIRPB2 |
| rs62078596 | chr17 | 53,906,496 | C/T | 8.50E-07 | 0.013 | [0.008,0.018] | PCTP--[]---ANKFN1 |
| rs8134126 | chr21 | 28,401,705 | C/T | 1.00E-06 | -0.01 | [-0.013,-0.006] | ADAMTS5--[] |
| rs138108783 | chr1 | 119,711,419 | A/G | 1.40E-06 | 0.274 | [0.163,0.385] | WARS2--[]---HAO2 |

Supplementary Table 5: Variants associated with hotspot usage. Linear regression model tested as hotspot_usage ~ sex + age + pc.0 + pc.1 + pc.2 + pc.3 + pc.4 + genotype. Association tests conducted using only individuals found to have ≥ 97% European ancestry.

| Population | Female sample size* | Male sample size* | Female median hotspot usage | Male median hotspot usage | Difference | p-value (Mann-Whitney U) |
|---|---|---|---|---|---|---|
| Europe | 3329 | 3325 | 62.96% | 67.12% | 4.16% | **4.93E-40** |
| Latino | 362 | 341 | 61.15% | 66.84% | 5.68% | **1.36E-09** |
| East Asia | 221 | 180 | 60.38% | 67.56% | 7.18% | **5.67E-06** |
| South Asia | 97 | 95 | 61.65% | 66.35% | 4.71% | **0.00494563** |
| Middle East | 88 | 88 | 59.52% | 61.26% | 1.74% | 0.284789 |
| African American | 43 | 57 | 61.37% | 65.37% | 4.00% | 0.135323 |
| **All** | **5668** | **5621** | **0.6268** | **0.67255** | **0.04575** | **1.06E-69** |

Supplementary Table 6: Differences in hotspot usage between males and females, partitioned by population. *The sample size represents the number estimated $\alpha$'s, with one estimate for each meiosis from phase-known parents, and a single estimate for phase-unknown parents.

| Females | | | | | | | | | |
|---|---|---|---|---|---|---|---|---|---|
| | Gamma Model (no escape) | | | Escape Model | | | | | |
| | Phase known | Phase unknown | Weighted mean | Phase known | | Phase unknown | | Weighted mean | |
| Chrom | $\nu$ | $\nu$ | $\nu$ | $\nu$ | p | $\nu$ | p | $\nu$ | p |
| chr1 | 2.749 | 3.211 | **2.952** | 6.045 | 0.067 | 6.711 | 0.079 | **6.384** | **0.073** |
| chr2 | 2.390 | 3.035 | **2.643** | 6.499 | 0.064 | 6.902 | 0.076 | **6.718** | **0.070** |
| chr3 | 2.328 | 2.653 | **2.473** | 6.489 | 0.072 | 6.612 | 0.089 | **6.556** | **0.081** |
| chr4 | 3.074 | 3.956 | **3.414** | 5.981 | 0.042 | 6.036 | 0.047 | **6.009** | **0.044** |
| chr5 | 3.289 | 3.824 | **3.526** | 6.582 | 0.044 | 6.941 | 0.065 | **6.753** | **0.052** |
| chr6 | 2.893 | 2.864 | **2.878** | 7.221 | 0.055 | 7.395 | 0.086 | **7.314** | **0.069** |
| chr7 | 3.007 | 2.826 | **2.902** | 7.435 | 0.048 | 7.289 | 0.090 | **7.360** | **0.065** |
| chr8 | 1.395 | 2.014 | **1.566** | 8.073 | 0.165 | 6.615 | 0.184 | **7.141** | **0.175** |
| chr9 | 1.760 | 2.590 | **2.007** | 6.168 | 0.095 | 7.096 | 0.113 | **6.586** | **0.105** |
| chr10 | 2.548 | 4.228 | **2.971** | 7.561 | 0.066 | 7.039 | 0.056 | **7.260** | **0.061** |
| chr11 | 2.485 | 2.829 | **2.645** | 7.466 | 0.065 | 8.240 | 0.084 | **7.818** | **0.074** |
| chr12 | 2.979 | 3.896 | **3.323** | 7.519 | 0.058 | 6.927 | 0.060 | **7.175** | **0.059** |
| chr13 | 3.506 | 4.727 | **3.982** | 7.876 | 0.039 | 7.157 | 0.034 | **7.442** | **0.036** |
| chr14 | 2.654 | 4.065 | **3.070** | 7.574 | 0.056 | 7.338 | 0.059 | **7.451** | **0.057** |
| chr15 | 2.090 | 2.604 | **2.292** | 7.652 | 0.081 | 7.842 | 0.109 | **7.754** | **0.095** |
| chr16 | 1.357 | 1.888 | **1.504** | 7.708 | 0.158 | 9.383 | 0.220 | **8.277** | **0.190** |
| chr17 | 2.874 | 4.016 | **3.246** | 8.216 | 0.064 | 6.972 | 0.056 | **7.479** | **0.061** |
| chr18 | 3.063 | 4.920 | **3.575** | 8.244 | 0.064 | 8.056 | 0.053 | **8.139** | **0.058** |
| chr19 | 3.444 | 5.322 | **4.001** | 7.991 | 0.052 | 8.576 | 0.055 | **8.273** | **0.053** |
| chr20 | 3.149 | 3.530 | **3.329** | 7.672 | 0.060 | 7.612 | 0.078 | **7.637** | **0.070** |
| chr21 | 2.694 | 3.596 | **2.996** | 9.454 | 0.061 | 9.713 | 0.064 | **9.598** | **0.062** |
| chr22 | 2.315 | 1.904 | **2.033** | 9.456 | 0.060 | 10.664 | 0.128 | **9.958** | **0.090** |
| chrX | 1.959 | 2.151 | **2.050** | 6.439 | 0.089 | 5.886 | 0.110 | **6.129** | **0.100** |
| Autosomes | 2.409 | 3.084 | **2.666** | 7.134 | 0.071 | 7.233 | 0.086 | **7.188** | **0.078** |

| Males | | | | | | | | | |
|---|---|---|---|---|---|---|---|---|---|
| | Gamma Model (no escape) | | | Escape Model | | | | | |
| | Phase known | Phase unknown | Weighted mean | Phase known | | Phase unknown | | Weighted mean | |
| Chrom | $\nu$ | $\nu$ | $\nu$ | $\nu$ | p | $\nu$ | p | $\nu$ | p |
| chr1 | 3.240 | 3.289 | **3.266** | 8.515 | 0.047 | 9.419 | 0.082 | **8.949** | **0.063** |
| chr2 | 4.081 | 3.972 | **4.019** | 7.567 | 0.038 | 8.439 | 0.063 | **8.024** | **0.050** |
| chr3 | 3.640 | 4.381 | **3.977** | 9.123 | 0.045 | 8.376 | 0.053 | **8.695** | **0.049** |
| chr4 | 4.469 | 4.256 | **4.343** | 8.516 | 0.046 | 9.217 | 0.072 | **8.895** | **0.059** |
| chr5 | 4.425 | 5.232 | **4.795** | 7.593 | 0.030 | 7.847 | 0.047 | **7.737** | **0.038** |
| chr6 | 3.255 | 3.388 | **3.324** | 9.828 | 0.055 | 9.199 | 0.077 | **9.456** | **0.066** |
| chr7 | 3.266 | 5.311 | **3.873** | 8.297 | 0.057 | 8.991 | 0.055 | **8.685** | **0.056** |
| chr8 | 2.197 | 1.816 | **1.946** | 10.760 | 0.119 | 9.216 | 0.173 | **9.775** | **0.145** |

| | | | | | | | | | | |
|---|---|---|---|---|---|---|---|---|---|---|
| chr9 | 2.137 | 3.642 | **2.490** | 9.253 | 0.108 | 9.845 | 0.096 | **9.587** | **0.101** |
| chr10 | 4.323 | 4.823 | **4.564** | 8.575 | 0.047 | 9.556 | 0.071 | **9.031** | **0.058** |
| chr11 | 3.693 | 4.879 | **4.160** | 7.422 | 0.055 | 8.794 | 0.058 | **8.158** | **0.057** |
| chr12 | 3.228 | 4.430 | **3.666** | 8.269 | 0.060 | 8.025 | 0.063 | **8.126** | **0.061** |
| chr13 | 5.706 | 4.058 | **4.467** | 8.387 | 0.029 | 10.051 | 0.058 | **9.142** | **0.042** |
| chr14 | 4.647 | 5.348 | **4.969** | 9.479 | 0.028 | 9.083 | 0.042 | **9.295** | **0.033** |
| chr15 | 2.579 | 3.596 | **2.932** | 8.127 | 0.065 | 9.244 | 0.064 | **8.652** | **0.064** |
| chr16 | 3.485 | 2.641 | **2.875** | 7.675 | 0.064 | 8.492 | 0.105 | **8.114** | **0.088** |
| chr17 | 3.278 | 2.092 | **2.339** | 8.735 | 0.063 | 9.582 | 0.125 | **9.220** | **0.095** |
| chr18 | 4.587 | 3.191 | **3.538** | 8.380 | 0.050 | 8.278 | 0.066 | **8.314** | **0.058** |
| chr19 | 3.808 | 4.607 | **4.156** | 7.423 | 0.061 | 8.975 | 0.074 | **8.104** | **0.068** |
| chr20 | 3.184 | 3.478 | **3.333** | 8.205 | 0.079 | 9.601 | 0.084 | **8.905** | **0.082** |
| chr21 | 2.485 | 5.772 | **2.841** | 100 | 0.074 | 100 | 0.049 | **100** | **0.057** |
| chr22 | 2.467 | 3.414 | **2.786** | 10.442 | 0.059 | 16.799 | 0.074 | **12.670** | **0.069** |
| **Autosomes** | 3.346 | 3.591 | **3.470** | 8.608 | 0.058 | 9.184 | 0.077 | **8.931** | **0.067** |

Supplementary Table 7: Interference parameter estimates for females (top) and males (bottom). Estimates are given for phase-known and phase-unknown individuals separately. In addition, a combined estimate was calculated as a weighted average with weights taken to be the reciprocal of the variance.

| Chrom | Start position (bp) | End position (bp) |
|---|---|---|
| 1 | 144,954,851 | 145,394,955 |
| 1 | 145,547,963 | 146,508,934 |
| 1 | 146,997,245 | 147,093,887 |
| 1 | 147,162,445 | 147,205,770 |
| 1 | 147,210,993 | 147,222,372 |
| 1 | 147,375,981 | 147,782,284 |
| 8 | 6,881,638 | 8,119,716 |
| 8 | 11,088,131 | 11,096,553 |
| 8 | 11,251,705 | 11,256,184 |
| 8 | 11,330,364 | 11,332,026 |
| 8 | 11,354,933 | 11,359,638 |
| 8 | 11,363,950 | 11,372,141 |
| 8 | 11,406,175 | 11,476,726 |
| 8 | 11,486,220 | 11,496,193 |
| 8 | 11,501,265 | 11,503,333 |
| 8 | 11,514,144 | 11,516,373 |
| 8 | 11,533,384 | 11,570,036 |
| 8 | 11,722,125 | 11,755,513 |
| 8 | 11,763,932 | 11,799,654 |
| 8 | 11,830,877 | 11,846,482 |
| 8 | 11,857,317 | 12,559,475 |
| 10 | 46,076,235 | 47,597,927 |
| 10 | 47,611,631 | 48,324,245 |
| 10 | 48,368,273 | 48,380,952 |
| 10 | 48,400,458 | 48,427,246 |
| 10 | 48,440,744 | 48,471,020 |
| 10 | 48,489,541 | 48,508,137 |
| 10 | 48,512,114 | 48,545,527 |
| 10 | 50,122,109 | 50,163,975 |
| 10 | 50,382,038 | 50,382,478 |
| 10 | 50,451,843 | 50,471,176 |
| 10 | 50,568,814 | 50,585,177 |
| 10 | 50,615,087 | 50,615,806 |
| 10 | 50,623,895 | 50,643,498 |
| 10 | 50,821,243 | 50,824,244 |
| 10 | 50,824,619 | 51,559,469 |
| 10 | 135,160,950 | 135,195,332 |
| 10 | 135,202,594 | 135,257,091 |
| 10 | 135,347,727 | 135,349,367 |
| 10 | 135,351,362 | 135,352,100 |
| 12 | 8,000,912 | 8,021,932 |
| 15 | 22,876,889 | 22,908,392 |
| 15 | 22,909,207 | 22,918,657 |
| 15 | 22,932,511 | 23,053,839 |
| 16 | 21,327,273 | 21,620,270 |
| 19 | 2,098,015 | 2,099,820 |
| 19 | 54,077,870 | 54,106,839 |
| 19 | 54,107,686 | 54,111,568 |
| 22 | 17,729,044 | 17,731,977 |
| 22 | 25,650,406 | 25,848,811 |

Supplementary Table 8: Locations of regions with high numbers of double recombination events. Hg19 coordinates.

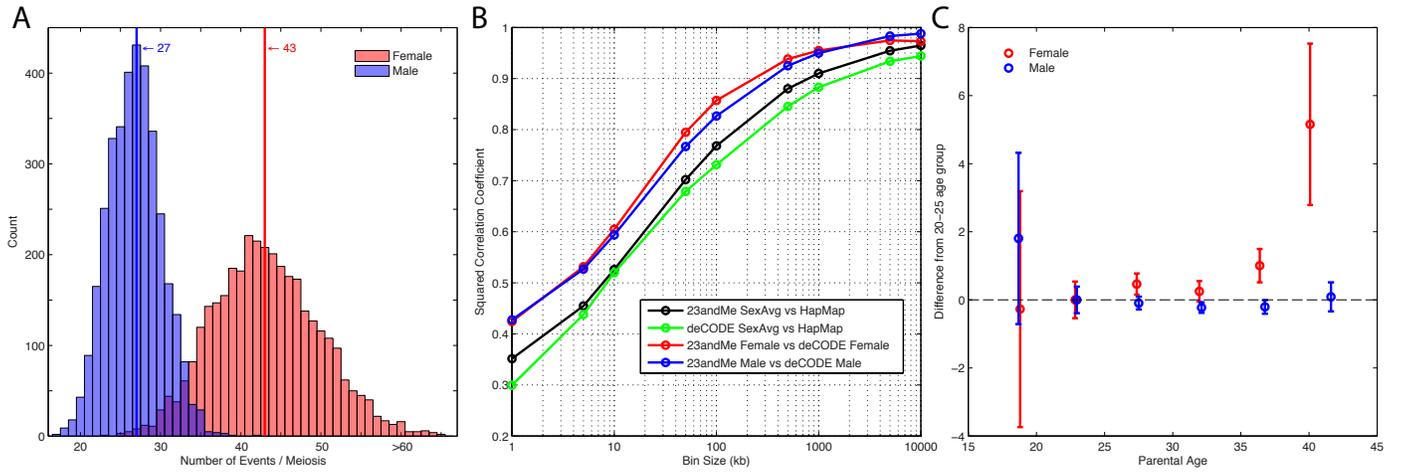

Figure 1

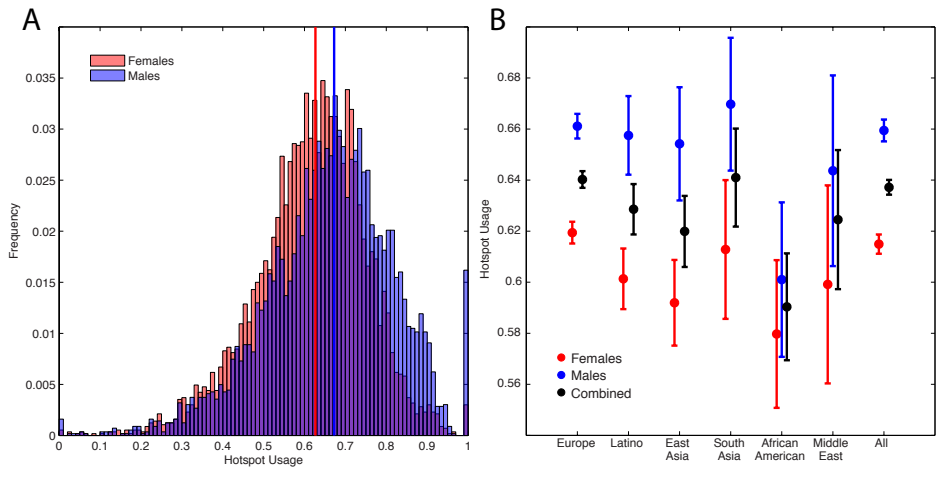

Figure 2

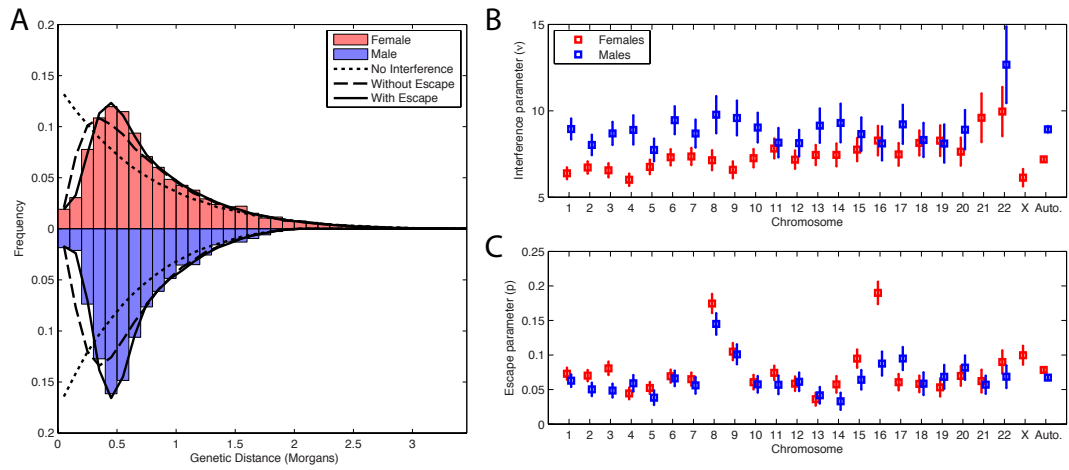

Figure 3

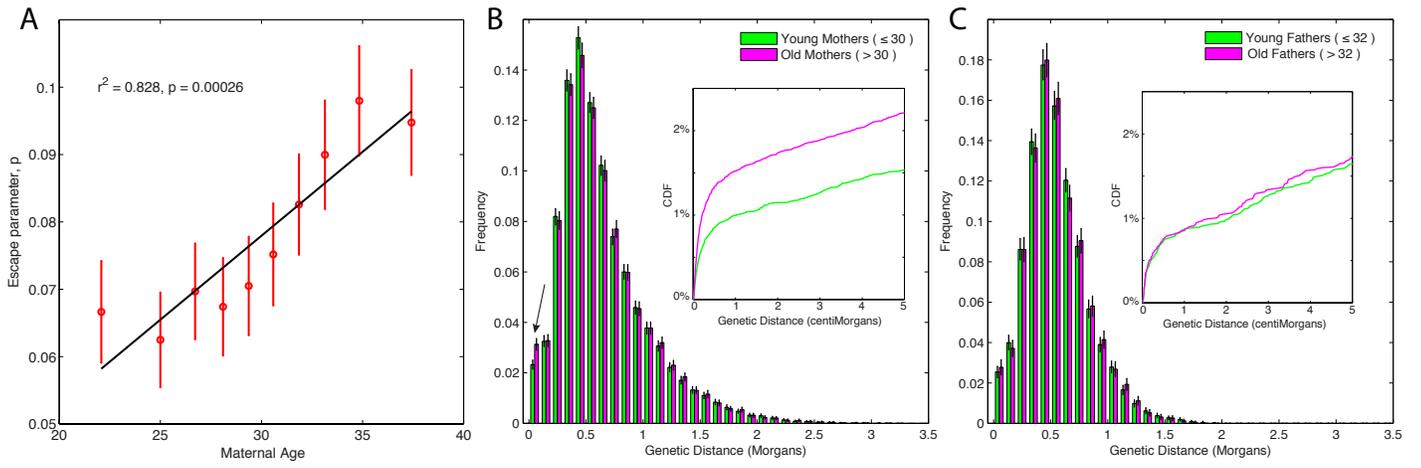

Figure 4

# Supplementary Material

## 1. Data Filtering

Prior to data filtering, the dataset consisted of 4,270 pedigree families, with data pertaining to 18,647 informative meioses. This raw dataset consisted of 692,876 recombination events, with a median of 45 and 28 events per meiosis in females and males respectively.

### 1.1. Step 1 – Removal of weakly supported events

The Merlin algorithm used to detect recombination events does not account for genotyping error, and genotyping errors are therefore likely to result in spurious recombination event calls. To account for this issue, our first step was to only use high-confidence sites. First, we required the sites to have a call rate greater than 90% and Hardy-Weinberg p-value $\leq 1 \times 10^{-20}$ (as calculated in the 23andMe cohort). Second, we excluded sites with minor allele frequencies differing from those of the 1000 Genomes Phase 1 reference panel (*21*). This was achieved by constructing a 2x2 contingency table and comparing the 1000 Genomes European allele counts with those from 2,000 randomly selection 23andMe customers, and using a chi-square test to identify significant deviations. Sites with p-values less than $1 \times 10^{-15}$ where removed.

Having applied these basic site filters, we next aimed to remove any weakly supported recombination events. This was achieved by first using the Merlin 'error' feature to remove potential genotyping errors not consistent with gene flow within each pedigree. In addition, we excluded all recombination events supported by less than three recombination-informative sites on either side, where we define an informative site as a site that is called as heterozygotic in exactly two individuals out of each mother-father-child trio. Finally, we removed all pairs of events within each single family that occurred within the same SNP interval. Together, these filters removed 31,742 weakly supported events, which corresponded to 4.6% of the total number.

### 1.2. Step 2 – Removal of clustered double events

Preliminary inspection of the genetic maps identified a region on chromosome 10 where the 23andMe genetic map diverged substantially from that generated by deCODE (*2*). This can be seen in a plot of the chromosome 10 genetic map at approximately 50Mb (Supplementary Figure 2A).

Further investigation of this region revealed a large number of 'double' crossovers in close proximity to each other (i.e. pairs of recombination events occurring in close proximity within the same individual). While some such observations are expected through the action of gene conversion, such strong clustering of these events is not expected biologically. Instead, we believe the result is suggestive of misplacement of polymorphisms, mis-assembly of one or more reference contigs in the hg19 reference genome, or of more complex types of error related to copy number polymorphism or array design. In any case, these double recombination events represent a form of error that needed to be eliminated.

To better quantify this issue, we identified all pairs of recombination events occurring within a single individual that were within 1Mb of each other. For each SNP in the genome, we estimated the number of these event pairs that span the SNP (Supplementary Figure 2B).

For the vast majority of the genome, there were very few such event pairs, and hence localized peaks likely represent data quality issues. We therefore identified all SNPs spanned by at least than 14 event pairs (with this threshold being equivalent to the 99.9th percentile of the distribution). In this way, we identified 50 regions with strong enrichment of nearby event pairs (Supplementary Table 8). Note that for this analysis we ignored the pseudoautosome, as a large number of events occurring in close proximity might be expected due to the extreme male recombination rate within this region.

The regions with high numbers of clustered events were themselves clustered into 13 regions across 8 chromosomes, and are often in the vicinity of chromosome centromeres, telomeres, or reference assembly gaps. We removed all event pairs within 500kb of the region boundaries described in Supplementary Table 8, which resulted in the removal of 2,916 events (0.42% of the total). The removal of these events improved the concordance between the 23andMe and deCODE maps (Supplementary Figure 2C).

### 1.3. Step 3 – Removal of individuals with unusual total numbers of events

Previous research using well-curated data in 728 meioses reported an average of 39.6 autosomal events per gamete in females (95% C.I. 38.5 – 40.6), and 26.2 autosomal events per gamete in males (95% C.I. 25.6 – 26.7) (*7*). The minimum/maximum number of observed autosomal events in any given meiosis in this data was 19/71 for females, and 16/43 for males (Graham Coop, personal communication).

Preliminary analysis of our data revealed a small subset of individuals had biologically unrealistic numbers of recombination events. Our first filtering step was to remove the pedigrees containing these individuals. Specifically, we removed individuals (and their containing pedigrees) that were more than 5 standard deviations from the (sex specific) median number of recombination events. To guard against outliers, we used a robust estimate of the standard deviation taken as $\sigma = 1.4826\ MAD$, where MAD represents the median absolute deviation.

Before filtering the median number of recombination events was 43 and 27 for females and males respectively (including chrX and the pseudoautosome). Using the $\pm 5\sigma$ thresholds, we removed pedigrees containing any female with fewer than 10 or more than 76 events per meiosis, or any male with fewer than 9 or more than 45 events. These filters removed a total of 52 pedigrees.

### 1.4. Summary of filtered dataset

After applying the filtering steps described above, the filtered dataset consists of 4,209 pedigrees containing 18,302 informative meioses, of which 9,152 are from females and 9,150 are from males. Of the families included in the study, 78.6% are family quartets, 14.3% are larger single-generation families, and 7.1% are 2-generation families (Supplementary Table 1).

Due to the structure of the pedigrees included in the study, certain recombination events can be identified as having occurred within a specific child, whereas others cannot. For example, in family quartets, it is generally unclear which child has the recombinant haplotype and we therefore refer to these events as 'phase-unknown'. Conversely, the child containing the recombinant haplotype can generally be identified in larger pedigree families in which the parental haplotype can be confidently phased, and we therefore refer to these events as 'phase-known'.

In total, 4,276 meioses are derived from phase-known individuals, whereas 14,026 are derived from phase-unknown individuals. Of the female meioses, 2,184 are derived from phase-known mothers and 6,968 are from phase-unknown mothers. Of the male meioses, 2,092 are derived from phase-known fathers, and 7,058 are derived from phase-unknown fathers.

Individuals were assigned to high-level population groups via comparison to a set of reference populations (see below). The majority of individuals in the dataset are of European descent, with approximately 78% of the meioses in the sample occurring within a European individual (Supplementary Table 2).

The parental age distribution for the filtered dataset is shown in Supplementary Figure 1. The mean age was 30 years for females, and 32 for males.

The final filtered dataset consists of 645,853 recombination events. Including the sex chromosomes, the mean number of recombination events was 43.47 for females ($\sigma = 6.64$, 95% C.I. 43.25 – 43.69), and 27.04 for males ($\sigma = 3.28$, 95% C.I. 26.94 – 27.16). For the autosomes alone, the mean number of recombination events was 41.64 for females ($\sigma = 6.34$, 95% C.I. 41.43 – 41.85) and 26.61 for males ($\sigma = 3.26$, 95% C.I. 26.51 – 26.73).

The distribution of interval sizes to which crossovers could be resolved (i.e., the distance between informative markers on either side of the recombination event) is given in Supplementary Figure 3. Crossovers could be resolved within a median distance of 28.2kb.

### 2. Assessment of robustness to genotyping error

In order to understand how our results could be influenced by genotyping error we simulated data for each of the pedigree structures contained within our data. To do this, we generated haplotypes for the founder individuals using the coalescent simulation software *ms* (*22*). Specifically, we generated 6 haplotypes (using: ms 6 1 -t 2189.781) and combined haplotypes at random to generate the genotypes of the founders. The population mutation rate was selected give an expected number of 5000 segregating sites. Children were then created by drawing haplotypes from each parent, and adding recombination as required.

To test MERLIN's ability to detect crossover events we placed one recombination event in the center of the sequence in one random parent, and passed this simulated pedigree data to MERLIN for haplotype analysis (option --best). This process is repeated to obtain 1000 total events per parent in each pedigree structure. Our results indicate that MERLIN is able to capture 99.6% of recombination events generated in this manner. The false negative calls resulted from low levels of heterozygosity (i.e. high relatedness) in the simulated haplotypes. The events placed in phase-known pedigrees were correctly assigned to the proper child in all cases. We repeated this simulation in the absence of any introduced recombination and find that in all cases, no events were called.

Estimates of the error rate of the Illumina HumanOmniExpress array used by 23andMe range from 0.01% (*23*) to 0.054% (*24*). To test for robustness of our results to genotyping error, we next simulated pedigrees without recombination, but with a single genotyping error introduced into one of the individuals by switching one of the alleles at the middle site in the sequence. This procedure was repeated 1000 times in each of the five pedigree structures in our dataset. We looked for any events called by MERLIN and recorded the position in the sequence and the number of informative sites to the left and right of the event.

We estimated the number of false recombination events as a function of genotyping error. Without any filtering (and without using MERLIN's error detection functionality), we find that MERLIN to be sensitive to genotyping error. For a dataset of our size and pedigree composition, a genotyping error rate of 0.001% would produce 15,000 false positive recombination events, rising to 150,000 for a 0.01% genotyping error rate. However, the filters applied in the real dataset are effective at removing these simple false positives. After requiring at least 3 informative sites on both sides of a recombination event, we estimate that a dataset of our size would contain 74 spurious events with a 0.001% genotyping error rate, 739 with a 0.01% genotyping error rate, and 7,386 with a 0.1% genotyping error rate.

Although the assumptions of this simulation study are quite simplistic, given our dataset contains over 645,000 events these results would suggest that less than 1% of the events represent false positives. In addition, we note that in analysis of the real data, we used high-confidence sites and removed potential genotyping errors using MERLIN's error-detection feature (see Supplementary Section 1.1).

### 3. Individual Ancestral Assignment

Individuals were assigned to ancestral categories by quantifying the genetic variation they share with a set of representative reference populations. Chromosomal segments are assigned to geographic regions using 23andMe's Ancestry Composition tool (*25*). Informally, Ancestry Composition assigns regions of an individual's genome to 31 reference populations constructed from public reference datasets as well as private 23andMe cohort data (*26*). Individuals are assigned to genomic regions by first splitting the genome into short non-overlapping segments, and assigning each segment to the reference population with the highest degree of similarity. Given this assignment, it is straightforward to compute the percentage of an individual's DNA that originates from a certain subpopulation. For example, if 200,000 out of 400,000 total segments are predicted to come from an African background, then the global percentage of African ancestry is 50%. Given this global percentage, individuals are assigned to high-level categories (European, Middle Eastern, East Asian or South Asian) if their total percentage of ancestry in that category exceeds 97%. For individuals of admixed ancestry, 23andMe uses a logistic classifier trained on the segment length distributions of individuals who have self-identified as African American or Latino.

In order to define the final population label for a given individual, we first determined if they had at least 97% European, Middle Eastern, East Asian or South Asian ancestry. If so, then their category was determined. If the 97% threshold was not met, but the individual had a total global percentage of at least 97% when summing contributions from European, African and Native American, then the logistic classifier was applied. If neither of these conditions were met, then the individual was categorized as 'Other'.

## 4. Estimation of hotspot usage

To estimate the degree of hotspot usage by an individual, we adopted the method of Coop *et al.* (*7*). In brief, this method estimates the fraction of recombination events that overlap with known LD-based hotspots while accounting for the uncertainty in the localization of the called recombination events. For convenience, we re-describe the approach here.

We aim to estimate the proportion, $\alpha$, of events that occur within LD-based hotspots. Given a recombination event, $r$, the probability that the event overlaps with a hotspot is given by:

$$P(r \text{ overlaps a hotspot}) = \alpha + (1 - \alpha)P(r \text{ overlaps a hotspot by chance})$$

To estimate $P(r \text{ overlaps a hotspot by chance})$, we randomly shift the recombination events by a normally distributed distance (mean 0, standard deviation 200kb) a total of 1,000 times, and calculated the fraction of these moves that result in the event overlapping a hotspot. The likelihood for $\alpha$ is given by:

$$L(\alpha|r) = \delta_r P(r \text{ overlaps a hotspot}) + (1 - \delta_r)\big(1 - P(r \text{ overlaps a hotspot})\big)$$

where $\delta_r$ is an indicator function, taking the value 1 if $r$ overlaps a hotspot and zero otherwise. For a set of $k$ recombination events labeled $r_0, r_1, \ldots, r_{k-1}$, the likelihood of $\alpha$ for the whole dataset is given by:

$$L(\alpha|r_0, r_1, \ldots, r_{k-1}) = \prod_{i=0}^{k-1} L(\alpha|r_i)$$

We used this method to estimate $\alpha$ for each mother and father (for phase unknown individuals), and each meiosis (for phase known individuals). As in Coop *et al.*, we used all events that were well localized to within 30kb, but note that our results are robust to larger values of this parameter. The likelihood of alpha was estimated over a uniformly spaced grid of 2,000 values between 0 and 1, with the MLE taken as the value of $\alpha$ with the maximum likelihood on this grid. A 95% confidence interval was constructed as being the set of values within two log likelihood units of the MLE.

For phase-known individuals for which recombination events could be assigned to specific children, a separate $\alpha$ was estimated for each meiosis. For phase-unknown individuals where such an assignment was not possible, $\alpha$ was estimated using all events that could be attributed to the parent.

### 4.1. Hotspot usage results

The estimates for hotspot usage are shown in Supplementary Figure 7. The median hotspot usage estimate for females was 62.68% (95% C.I. 62.25% - 63.10%), whereas for males it was 67.26% (95% C.I. 66.85% - 67.69%), a difference of 4.6% ($p = 1.1 \times 10^{-69}$, Mann-Whitney U).

To ensure the difference between males and females is not driven by higher precision in females (resulting from higher numbers of events), we thinned the female data in order to match the number of events in males. Specifically, for each male, we randomly selected a female (without replacement) with a greater or equal number of events, and thinned the female events to match the number of male events. The resulting dataset contains an equal number of males and females, with each pair having an equal number of events. The estimates of hotspot usage for the two sexes were very similar to the previous estimates (62.2% for females, and 66.8% for males), and the difference in hotspot usage remains highly significant ($p < 2.2 \times 10^{-16}$).

To determine whether the observed differences in hotspot usage between males and females is dependent on the position within the chromosome (as males tend to have higher recombination rates towards the telomeres), we repeated the analysis having divided each chromosome into segments. Specifically, we split each chromosome into three windows, assigning the terminal 25% of sequence from each end to p- and q-arm bins, and keeping the central 50% of the sequence for the middle bin. For acrocentric chromosomes we omit the p-arm bin. We estimated the degree of hotspot usage in each of these bins. We observe that males use hotspots to a greater extent than females (Mann-Whitney U $p < 2.2 \times 10^{-16}$ for all three bins), suggesting that the difference in hotspot usage between males and females cannot be explained by telomere effects.

Due to variation in PRDM9, hotspot usage is expected to vary between populations (*10, 11*). The hotspots used in this study were identified from genome-wide Phase II HapMap linkage disequilibrium data (*27*), in which hotspots were called that were active in at least two of the three constituent populations (CEU, YRI, JPT+CHB). As such, one possibility for the observed difference between males and females is that the ancestry proportions within our data differ between the female and male samples. Inspection of the ancestry proportions within our data showed this not to be the case. In addition, if the analysis is partitioned by inferred ancestry, females have lower hotspot usage within all populations (Figure 2B), with the difference remaining significant in European, East Asian, Latino, and South Asian populations (Supplementary Table 6).

## 5. Description of age effect

Previous research has indicated a relationship between maternal age and the number of recombination events. In particular, research from the deCODE consortium used data from 14,140 meioses to report that the number of recombination events in females increase with age (*6*). The reported effect size is reasonably modest, contributing 0.082 ($\pm$ 0.012 standard error). recombination events per year, depending on the analysis method used. This translates as approximately a 4% increase in the average maternal recombination rate over a period of 25 years. No such association was observed in males.

A second study confirmed this effect using 728 meioses observed with from Hutterite families (*7*), observing that mothers over 35 years of age had approximately 3.1 extra recombination events compared to those under 25. Despite the small sample size, the effect size in this study was estimated to be 0.19 ($\pm$ 0.092 standard error) events per year. Again, no such effect was observed in males.

Conversely, a separate research group considering recombination events in 195 meioses reported a decrease in the number of recombination events with maternal age (*28*). In this case, the effect size was larger, corresponding to between -0.49 and -0.42 crossovers per year, again with no such effect observed in males. Although the smallest of the three studies, the authors suggest that the discrepancy in the direction of the effect between studies could be due to marker density and/or true biological differences between populations.

### 5.1. Correlation between number of recombination and parental age

To quantify the correlation between parental age and recombination rate, we first partitioned our data into phase-unknown parents for which recombination events could not be assigned to a specific child (or meiosis), and phase-known parents for which such an assignment was possible. For the phase-unknown parents group we used the maternal / paternal ages averaged across children, whereas for the phase-known group, we used the known parental ages at the time of the child's birth.

Using linear regression, we estimated the association between the number of autosomal events and parent age (Supplementary Figure 6). A weak positive association between age and the number of recombination events was detected for females, but no such effect was observed for males. The number of recombination events in females increased on average by 0.067 per year (standard error: $\pm$ 0.0215), which is similar to the estimate from deCODE.

We note that the observed effect is quite weak, and appears to be largely driven by an increase in the number of recombination events for mothers of 35 years or older (Figure 1C).

To ensure the observed effect is not confounded by population structure within the data, we first repeated the analysis for each population separately. In Europeans, for whom we have by far the largest sample size (accounting for ~76% of individuals), a significant association with maternal age was still observed (0.087 extra events per year, $p = 3.2 \times 10^{-4}$). In all other populations (East Asian, Middle

Eastern, Latino, African American, and South Asian), no significant association was observed, possibly due to insufficient power. No significant association with paternal age was observed within any population.

## 6. Inferring Crossover Interference

In the following text, we provide a description of the crossover interference models used within the main analysis.

### 6.1. The Gamma Model (a.k.a. the 'simple interference' model)

We follow the description of the Gamma model of crossover interference presented by Broman and Weber (*13*). For clarity, we repeat the description of this the model below.

The Gamma model describes the locations of chiasmata on the four-strand bundle according to a stationary renewal process, with increments being drawn from a gamma distribution with shape $\nu$ and rate $2\nu$. As such, in this model the distances between chiasmata are independent with mean 0.5 Morgans, and a standard deviation of $(2\sqrt{\nu})^{-1}$. Under the assumption of no chromatid interference, the chiasmata are thinned such that each chiasmata becomes a crossover with probability 0.5. As such, this model satisfies the requirement that the average inter-crossover distance should be 1 Morgan.

The parameter $\nu$ is a unitless measure of the strength of interference. Specifically, $\nu = 1$ corresponds to no interference between chiasmata, and $\nu > 1$ corresponds to positive interference (i.e. decreased variance in chiasmata spacing than would be expected under a Poisson model), and $\nu < 1$ corresponds to negative interference (i.e. increased variance in chiasmata spacing than expected under a Poisson model).

Let $x_0, x_1, x_2, \ldots$ be the genetic distances (in Morgans) between adjacent chiasmata, with $x_0$ being the distance from the p-terminal end of the chromosome to the first chiasma. Under the gamma model, the chiasmata locations are generated according to a gamma renewal process, such that $x_1, x_2, \ldots$ are independent and follow a gamma distribution with shape $\nu$ and rate $2\nu$, where $\nu$ is a positive real number. Therefore, the density of $x_i$ is given by $f(x; \nu) = (2\nu)^\nu e^{-2\nu x} x^{\nu-1}/\Gamma(\nu)$, for $i > 0$, and where $\Gamma(.)$ represents the gamma function. The density of $x_0$ is given by $g(x; \nu) = 2[1 - F(x; \nu)]$, where $F$ is the cumulative distribution function (cdf) of $f$.

However, using transmitted genotype data, the actual chiasmata locations are not observed. Rather, only the crossovers derived from the chiasmata positions are observed. Assuming no chromatid interference, the probability that a chiasmata results in a crossover is ½.

Let $y_0, y_1, y_2, \ldots$ be the genetic distances (in Morgans) between adjacent crossovers. Each $y_i$ is independent, with density given by $f^*(y; \nu) = \sum_{k=1}^{\infty} \left(\frac{1}{2}\right)^k f_k(y; \nu)$, where is the gamma distribution density with shape $k\nu$ and rate $2\nu$: $f_k(k; \nu) = (2\nu)^{k\nu} e^{-2\nu x} x^{k\nu-1}/\Gamma(k\nu)$, which is derived from the convolution of $f(y; \nu)$ with itself $k$ times. The density of $y_0$ is given by

$g^*(k; v) = 1 - F^*(y; v)$, where $F^*$ is the cdf of $f^*$. Likewise, let $G^*$ represent the cdf of $g^*$.

Given the above model, the contribution to the likelihood is:

$$Lk(v; \mathbf{y}) = \begin{cases} 1 - G^*(L; v) & \text{if } m_i = 0 \\ g^*(y_0; v)g^*(y_1; v) & \text{if } m_i = 1 \\ g^*(y_0; v)\left[\prod_{j=1}^{m_i-1} f^*(y_j; v)\right] g^*(y_m; v) & \text{otherwise.} \end{cases}$$

The likelihood for the complete data may be obtained as the product over all individual contributions.

### 6.2. The Housworth-Stahl 'interference escape' model

The Gamma model assumes that all crossover events are subject to the same interference process. The model has been shown to fit the data reasonably well for numerous organisms(13, 29). However, evidence from model organisms suggests the existence of a subset of events that are not subject to crossover interference(30), and statistical support of this finding has been seen in humans(14, 15).

For this reason, we adopt the Housworth-Stahl model of interference, which models the distances between crossovers as being a mixture of two processes. In one process, crossovers are distributed according to the gamma model described above, whereas in the second process, crossovers are distributed without interference. We describe this model here, following Housworth and Stahl's 2003 paper(14), and refer to it as the 'interference escape' model.

Assume that we have a mixture of two independent types of crossover, such that one type occurs with probability $q$ and has interference parameter $v$, and the other type occurs with probability $p = 1 - q$ and is not subject to interference ($v = 1$). As for the Gamma model described above, let $x_0, x_1, x_2, \ldots$ be the genetic distances (in Morgans) between adjacent chiasmata, with $x_0$ being the distance from the p-terminal end of the chromosome to the first chiasma. The distances between chiasmata are given by a gamma distribution with shape $v$ and rate $2qv$. As such, the density of $x_i$ is given by $f(x; v, 2qv) = (2qv)^v e^{-2qvx} x^{v-1}/\Gamma(v)$, for $i > 0$. Likewise, the density of $x_0$ is given by $g(x; v, q) = 2q[1 - F(x; v, 2qv)]$, where $F$ is the cumulative distribution function (cdf) of $f$.

As described for the Gamma model, crossover events are determined by thinning the chiasmata positions, with each position retained with probability ½. Let $y_0, y_1, y_2, \ldots$ be the genetic distances (in Morgans) between adjacent crossovers of this type. Each $y_i$ is independent, with density given by $f^*(y; v, q) = \sum_{k=1}^{\infty} \left(\frac{1}{2}\right)^k f(y; kv, 2qv)$. The density of $y_0$ is given by $g^*(k; v, q) = q[1 - F^*(y; v, q)]$, where $F^*$ is the cdf of $f^*$. Likewise, let $G^*$ represent the cdf of $g^*$.

Now consider a dataset from a single meiosis where the intercrossover distances are given by $x_0, x_1, x_2, \ldots, x_n$, where $\sum_{i=0}^{n} x_i = L$. We assume these events are derived from two types of crossover. The interference-free type occurs with probability $p$ and has $v = 1$. The second type is subject to interference and occurs with probability $q = 1 - p$. To calculate the likelihood of the data, we must sum over the $2^n$ possible ways to assign crossovers to the two types. Given one possible assignment, we split the data into two sets of intercrossover distances, $y_0, y_1, y_2, \ldots, y_j$ for the interference-free type, and $z_0, z_1, z_2, \ldots, z_k$ for the second 'interference' type, where $j + k = n + 1$. The likelihood of the data in from the interference-free type is:

$$Lk(v = 1, q = p; \boldsymbol{y}) = \begin{cases} 1 - G^*(L; 1, p) & \text{if } j = 0 \\ g^*(y_0; 1, p)[1 - F^*(y_1|1, p)] & \text{if } j = 1 \\ g^*(y_0; 1, p)\left[\prod_{i=1}^{j-1} f^*(y_i; 1, p)\right][1 - F^*(y_j|1, p)] & \text{otherwise.} \end{cases}$$

The likelihood of the data from the interference type is:

$$Lk(v = t, q = 1 - p; \boldsymbol{z})$$
$$= \begin{cases} 1 - G^*(L; t, 1 - p) & \text{if } j = 0 \\ g^*(y_0; t, 1-)[1 - F^*(y_1|t, 1 - p)] & \text{if } j = 1 \\ g^*(y_0; t, 1 - p)\left[\prod_{i=1}^{j-1} f^*(y_i; t, 1 - p)\right][1 - F^*(y_j|t, 1 - p)] & \text{otherwise.} \end{cases}$$

To calculate the likelihood of the data, we sum over all $2^n$ possible assignments to the two types:

$$Lk'(v = t, q = p; \boldsymbol{x}) = \sum_{\substack{(y_0, y_1, y_2, \ldots, y_j), \\ (z_0, z_1, z_2, \ldots, z_k)}} Lk(v = 1, q = p; \boldsymbol{y}) \, Lk(v = t, q = 1 - p; \boldsymbol{z})$$

To calculate the likelihood over multiple individuals, one simply takes the product of the above likelihood.

In our implementation of the above formulas, we calculated $f^*$ by summing over $k$ from 0 to 25. Numerical integration was used to calculate $G^*$ using the *integral* function in *MATLAB*.

### 6.2.1. Extension to interference escape model for phase-unknown data

The above description of the interference escape model assumes that the observed crossover events can be assigned to a specific meiosis. However, in the case of the phase-unknown individuals that make up the majority of our data, the observed crossovers cannot be assigned to specific children. As such, the above model cannot be used.

To extend the model for phase-unknown, we perform the same trick of summing over all possible assignments to each type, but this time also summing over all possible assignments to each meiosis. Although this procedure is somewhat naïve, both simulations and comparison of results between phased and unphased families have shown that it works well in practice (Supplementary Figure 11).

Consider a family quartet. For each parent, the observed crossovers are the result of two independent meioses, which we will call $M_1$ and $M_2$ respectively. Let the intercrossover distances events in $M_1$ be $a_{i0}, a_{i1}, a_{i2}, ...$, and the intercrossover distances in $M_2$ be $b_{i0}, b_{i1}, b_{i2}, ....$, where $a_{i0}$ and $b_{i0}$ represent the distances between the first event and the p-terminal end of the chromosome in $M_1$ and $M_2$ respectively. If we could observe these intercrossover distances, we could apply the Housworth-Stahl model as described above. However, due to the nature of phase-unknown individuals, all we are unable to directly observe these distances, and can only observe crossovers derived from both meioses without knowing which event is from which meiosis.

Naively, we could be to sum over all possible assignments to each meiosis, and for each assignment apply the Housworth-Stahl model independently. However, this would be inefficient, as it would result in summing over $4^n$ possible assignments (as there are 2 crossover types in each of 2 meioses). Instead, we note that the same result can be achieved by combining the 'interference free' classes, allowing us to sum over $3^n$ possible assignments.

Let the $n$ observed crossover positions assigned to a parent be $z_{i1}, z_{i2}, z_{i3}, ..., z_{in}$, which are derived from a superposition of the gamma renewal processes. In order to calculate the likelihood of this data, we treat the assignment of each event as either belonging to one of two inference classes, or to a single interference free class. Specifically, we calculate the likelihood as:

$$Lk'(\nu, q; \mathbf{z}) = \sum_{\substack{(a_0, a_1, ..., a_i), \\ (b_0, b_1, ..., b_j), \\ (c_0, c_1, ..., c_k)}} [Lk(\nu = t, q = 1 - p; \mathbf{a}) Lk(\nu = t, q = 1 - p; \mathbf{b}) Lk(\nu = 1, q = 2p; \mathbf{b})]$$

where the summation is taken over all possible $3^n$ divisions of the $n$ crossovers into the three classes. The likelihood for the complete dataset is given by taking the product of $Lk'(\nu, q; \mathbf{z})$ over all individuals.

Maximum likelihood estimation of $\nu$ and $q$ was performed using a MATLAB implementation of the Nelder-Mead method (*31*), restricting the search space such that $\nu \in [0.1, 100]$, and $q \in (0, 0.5)$. Uncertainty in the MLE point estimates was obtained by using the inverse of the Fisher information matrix to estimate the covariance matrix.

We note that the mixture model lacks identifiability when $\nu$ is close to 1. In this situation, the estimates of $q$ become uninformative. When performing likelihood maximization, we experimented with including a weakly informative prior on $q$ that favors smaller values. Specifically, we set $P(q) = 1 - q$, and performing maximum *a posteriori* estimation in place of maximum likelihood. In simulations, we found this method slightly improved results when $\nu$ is small, and has negligible effect otherwise. However, given the limited benefit of this approach, we did not pursue it further.

We validated the extension using simulations, and found it to give comparable results to those obtained from the original version for phase-known data. In addition, the per-chromosome estimates obtained from the real data were largely concordant between the two estimates (Supplementary Table 7).

### 6.3. Interference across the genome

We fitted the Gamma and interference-escape models for each chromosome separately, and also having combined data across the autosomes. As reported previously (14, 15), we find the interference escape model to provide a much better fit to the data than the traditional Gamma model (Figure 3A), and therefore focus on parameter estimates from this model.

Across the whole genome, crossover interference is stronger in males than for females. The average interference parameter was estimated to be $v = 7.18$ in females, and $v = 8.93$ in males, which implies increased variance in crossover spacing for females relative to males. We infer that $p = 7.8\%$ and $p = 6.7\%$ of events escape interference in males and females respectively. We note that these estimates are quite similar to those obtained in Hutterites (15), where the estimates were reported as $v = 9.17, p = 8\%$ and $v = 6.96, p = 6\%$ in males and females respectively.

The results for each chromosome are shown in Figure 3B and C. In females, there is a clear trend of shorter chromosomes having higher interference parameter ($v$) estimates, whereas any such effect is much weaker in males. In contrast, no such relationship is seen in the fraction of events that escape interference ($p$).

Of note in males, the estimate of the interference parameter for chromosome 21 appears to be extremely large, if not infinite (Supplementary Table 7). This finding has been reported previously (13, 15), and reflects the fact that very few paternal chromosomes exhibit more than one crossover. In our data, just 1.7% of paternal meioses have evidence of more than one crossover on chromosome 21, compared to 30.0% for chromosome 20 and 8.3% for chromosome 22.

The degree of interference on a chromosome is reasonably well predicted by the map length. Combining data across the sexes, the chromosome map length explains 57% of the variance in the interference parameter (Supplementary Figure 8). When considering the sexes separately, the association is stronger in females (where 69% of the variance can be explained) than in males (where just 17.2% can be explained, and the fit does not achieve significance; $p = 0.061$). A multiple regression including sex as a predictor variable ($v_{chr} = \beta_0 + \beta_1 maplength_{chr} + \beta_2 sex_{chr}$) finds the $\beta_2$ to be marginally significant ($p = 0.0183$), but the model is not a significantly better fit than the model without including sex ($\Delta(\text{deviance}) = 3.54, p = 0.0599$).

### 6.4. Analysis of interference by age

We divided our data into quantiles of approximate equal size on the basis of age. For each decile, we fitted the interference-escape model. The results are shown in Figure 4 and Supplementary Figure 9 for 10 quantiles, and in Supplementary Figure 10 for 5 and 20 quantiles. In females, the proportion of events escaping interference consistently increases with maternal age, and the pattern is consistent across both phase-known and phase-unknown individuals (Supplementary Figure 11). There is no such correlation in the degree of interference, which appears to be constant across maternal ages. In contrast, no correlation is observed between paternal age and either parameter.

### 6.5. Stratified sampling to account for number of crossovers

One potential concern is that the inferred degree of interference may be influenced by a change in recombination rate. As the distribution of distances between crossovers depends on the number of crossovers (when there are more crossovers, they are necessarily more closely spaced), if there is a change in the recombination rate with age then this may influence the interference estimates.

We can address this concern by the use of stratified sampling. Specifically, for each age group, we subsampled individuals in order to ensure that each decile has the exact same distribution of the number of crossovers per meiosis. This was achieved as follows. First, for each age group $i$, we counted the number of individuals with $x$ crossovers, which we call $N_i(x)$. For each $x$, we estimated the minimum $N_i(x)$ across all decile age groups, so that $n(x) = \min_i(N_i(x))$. We then subsampled individuals within each decile by randomly selecting $n(x)$ individuals, without replacement, for each possible $x$.

Having performed this subsampling, we repeated the analysis. The results are shown in Supplementary Figure 12. The results for females are largely identical to that obtained without stratified sampling, with a significant increase in the proportion of events escaping interference as maternal age increases.

## Supplementary References

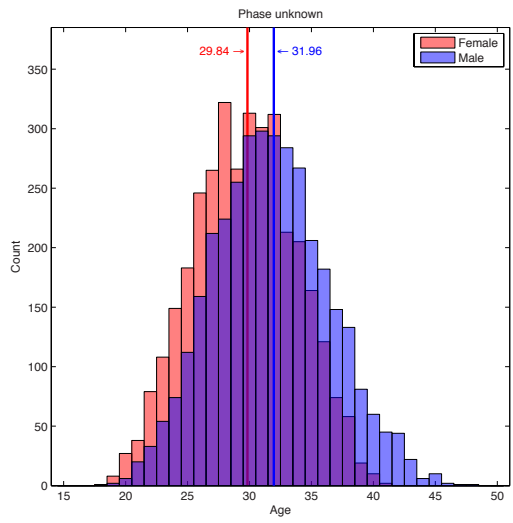 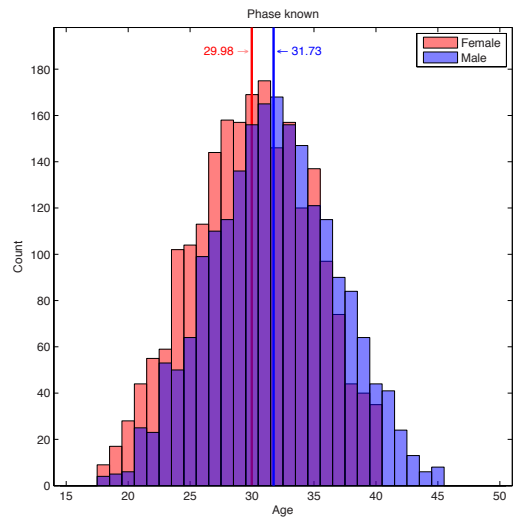

Figure S1

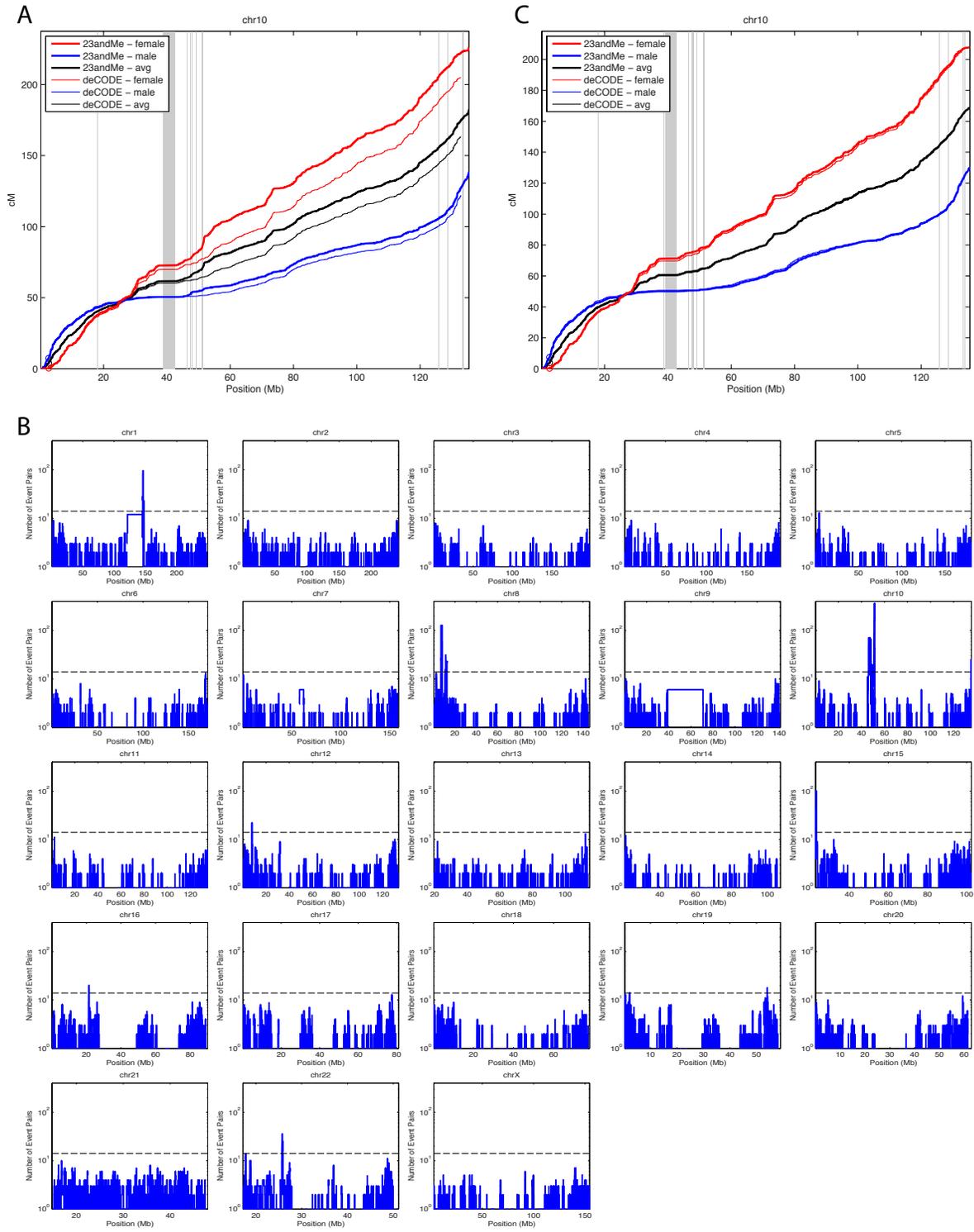

Figure S2

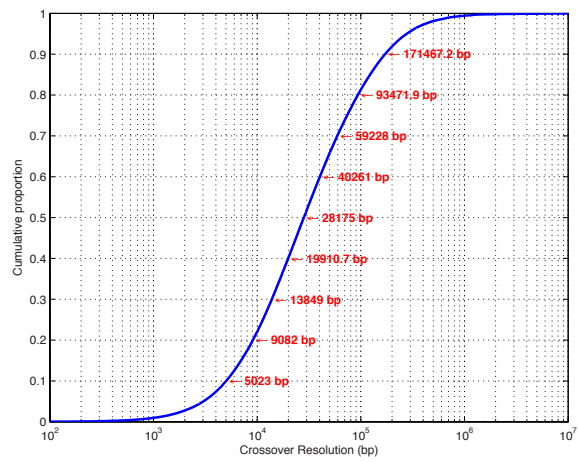

Figure S3

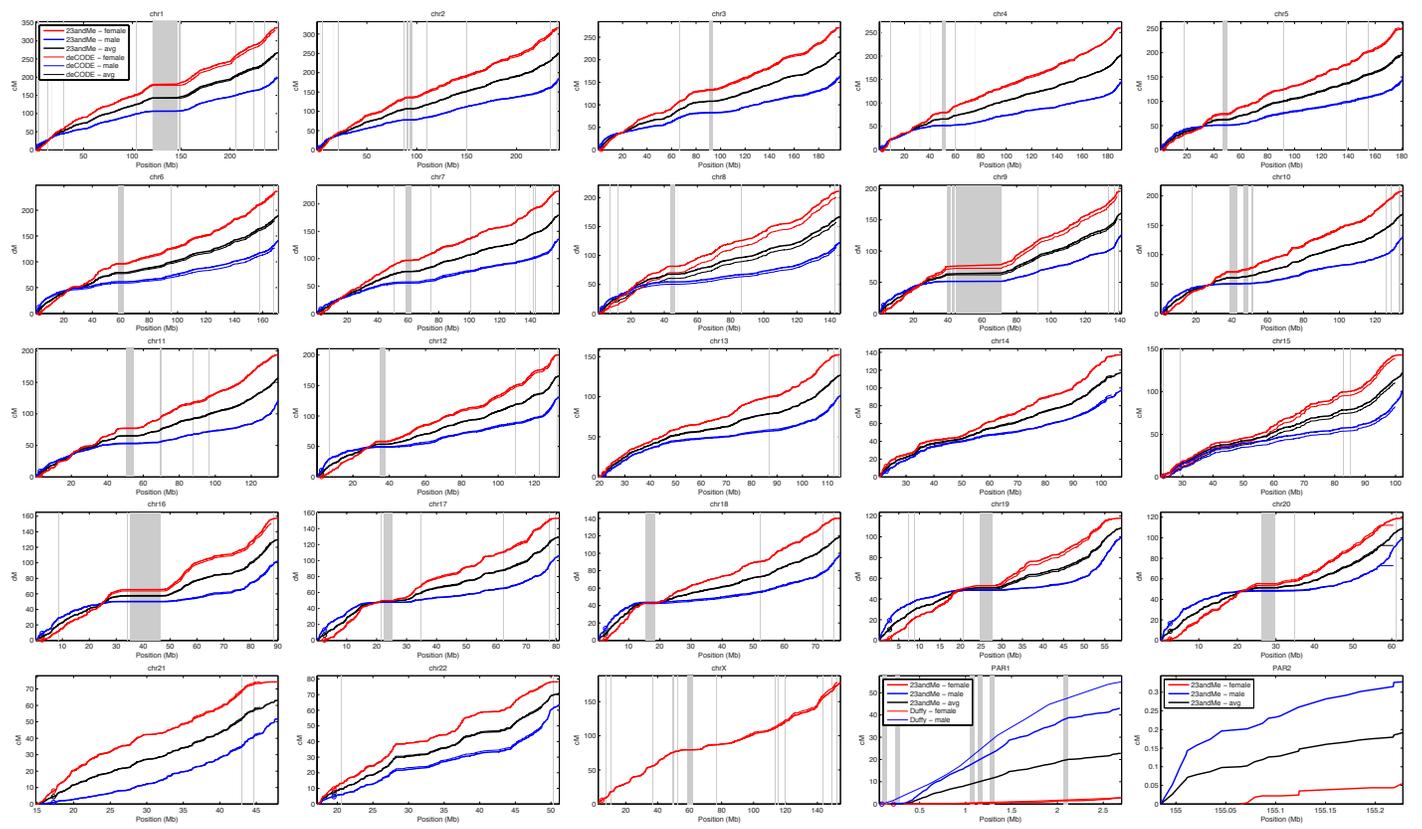

Figure S4

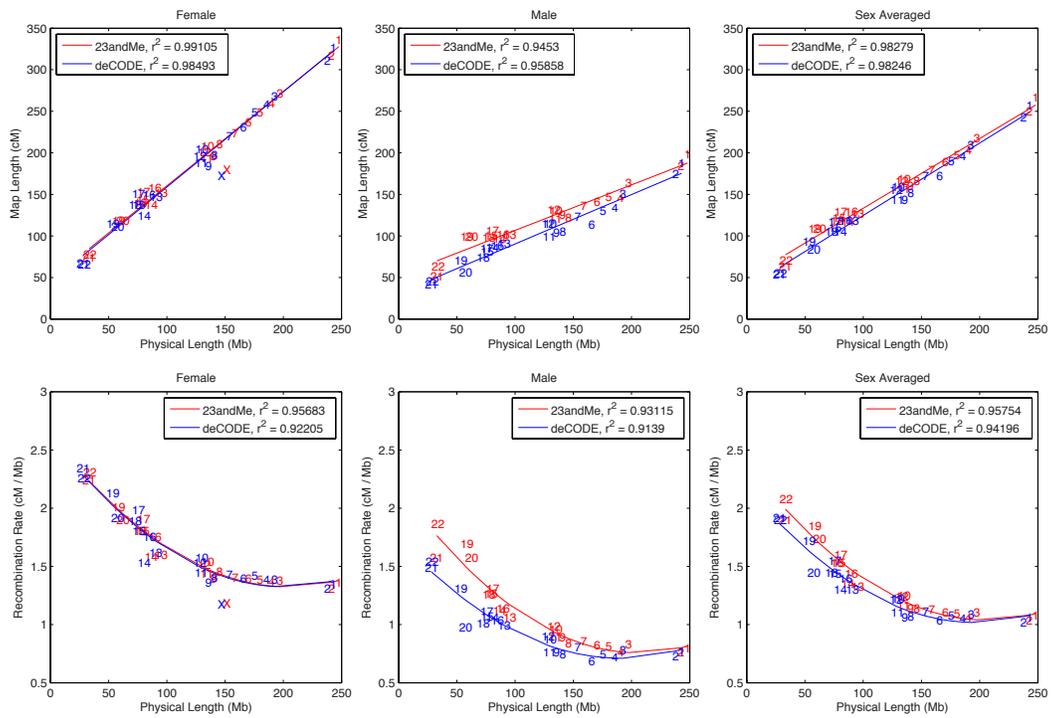

Figure S5

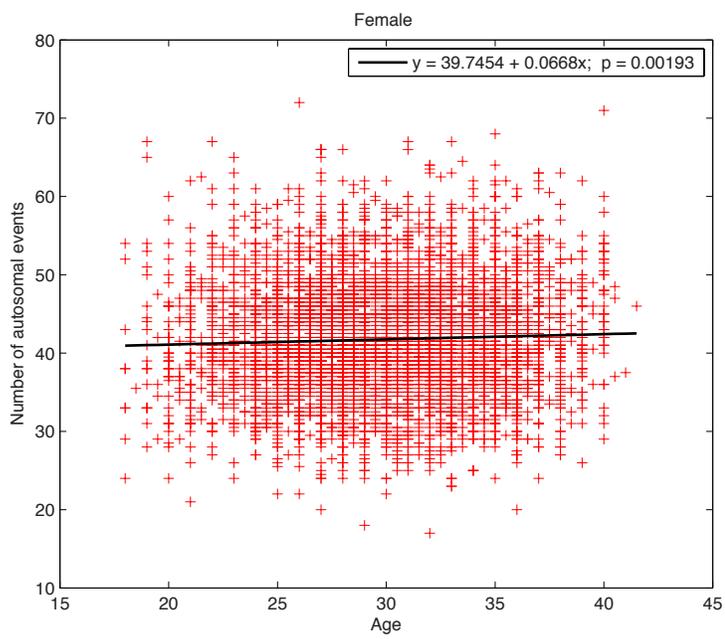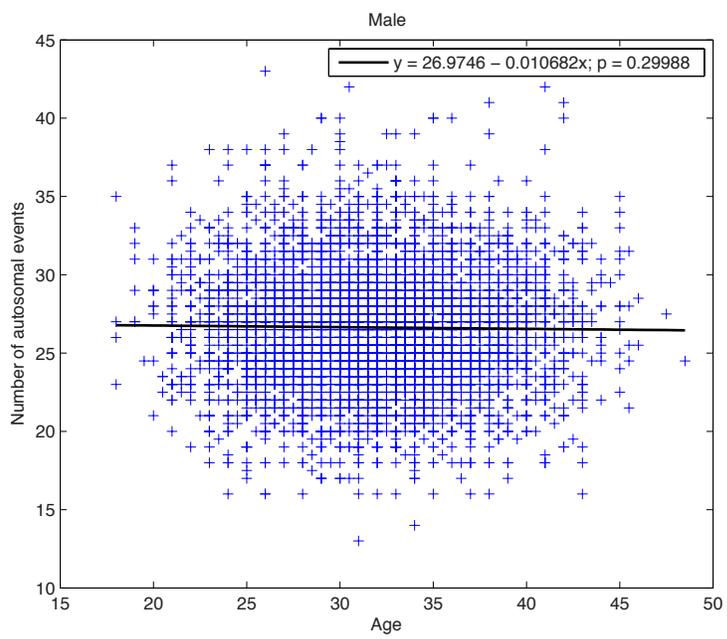

Figure S6

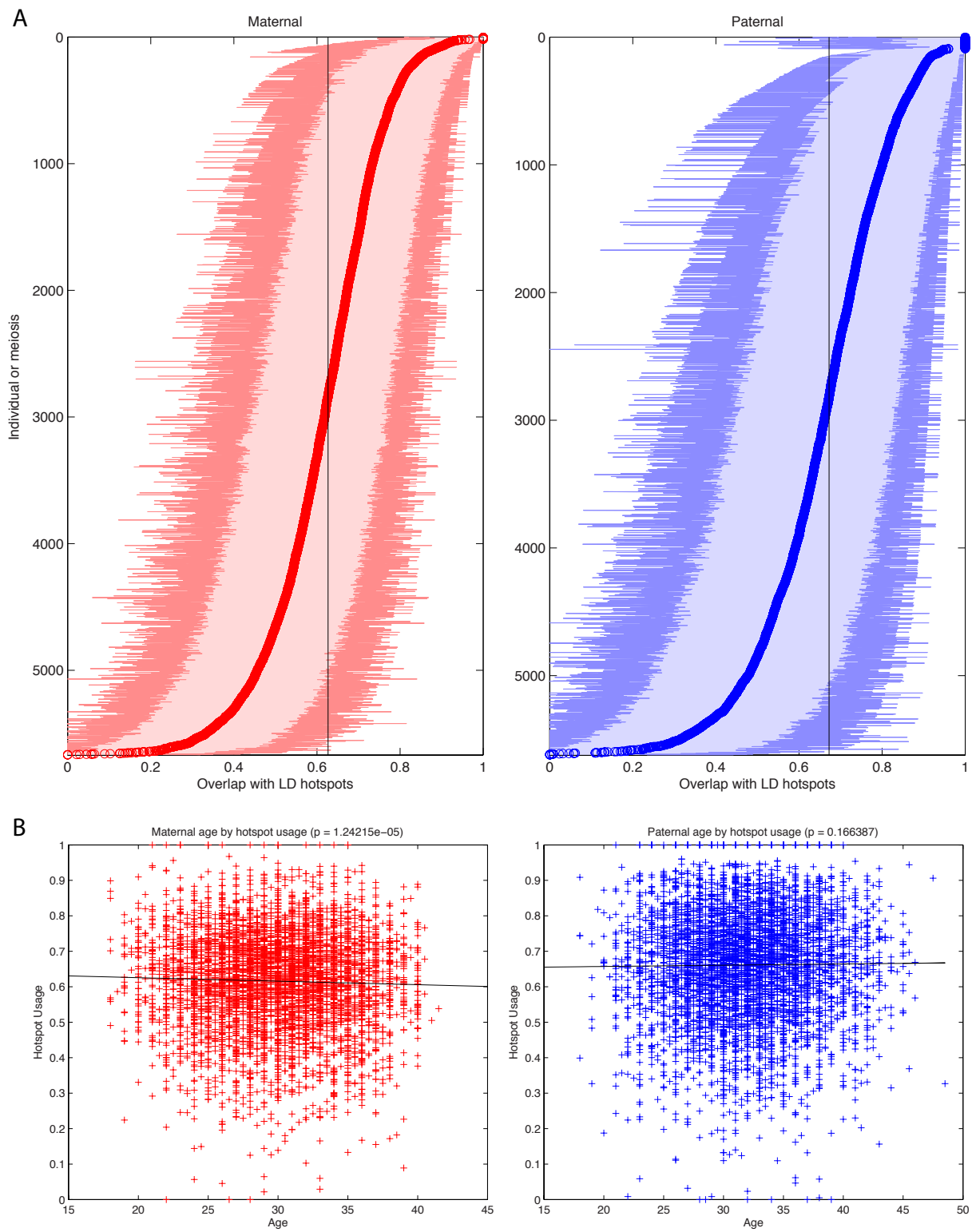

Figure S7

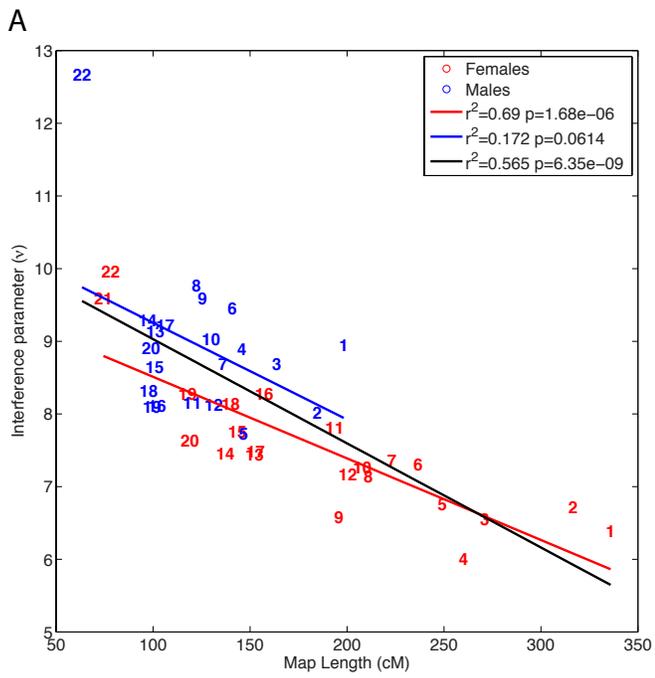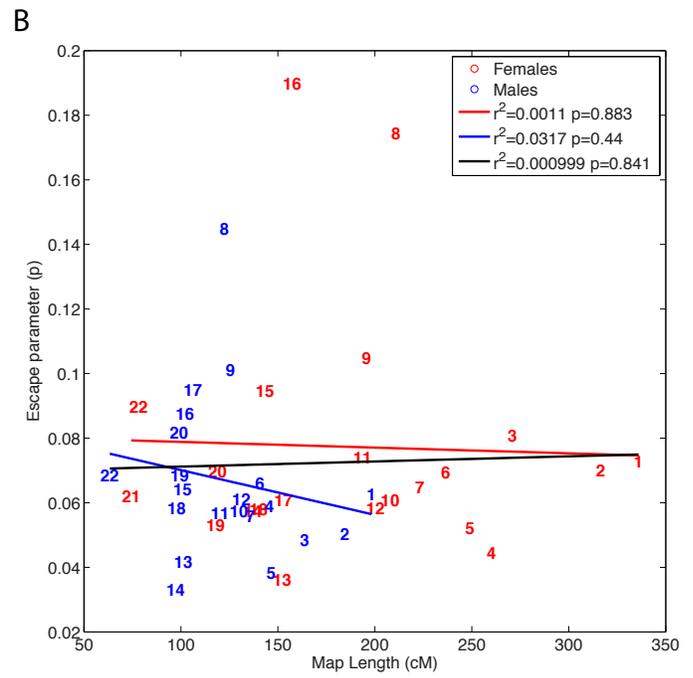

Figure S8

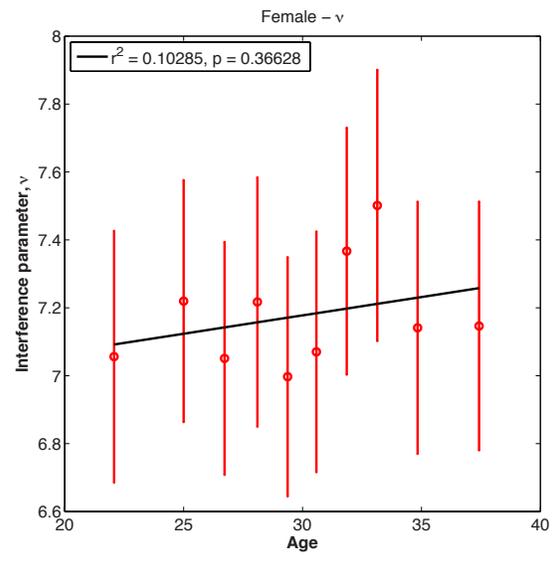
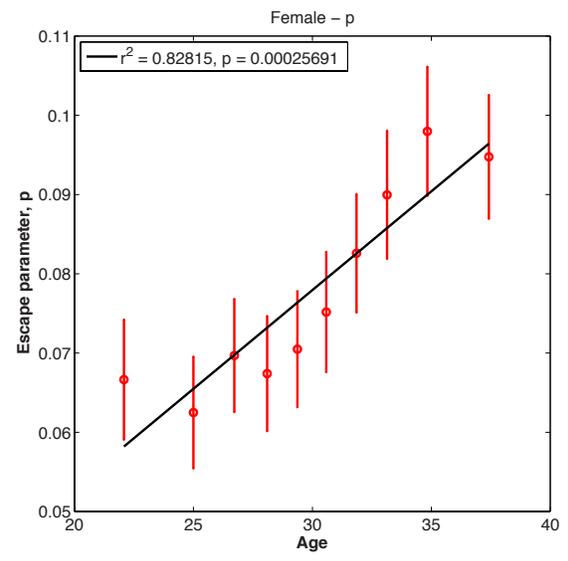
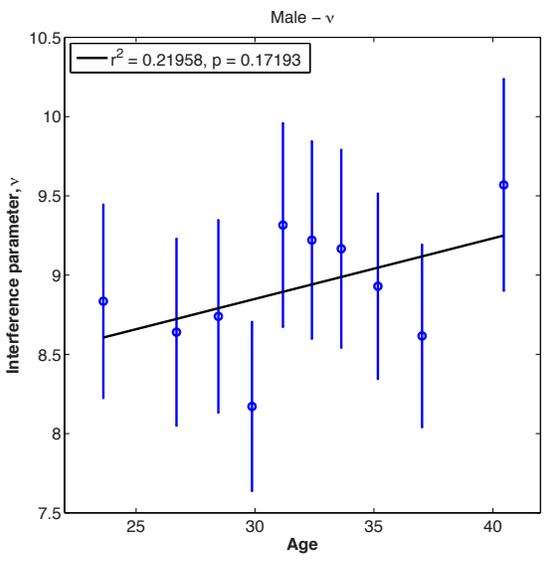
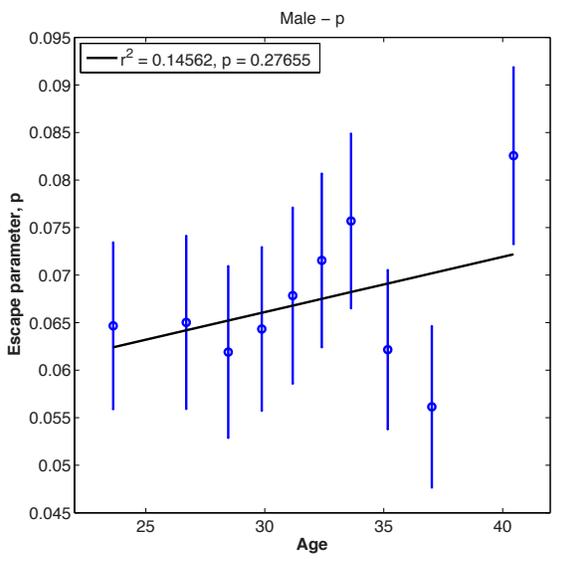

Figure S9

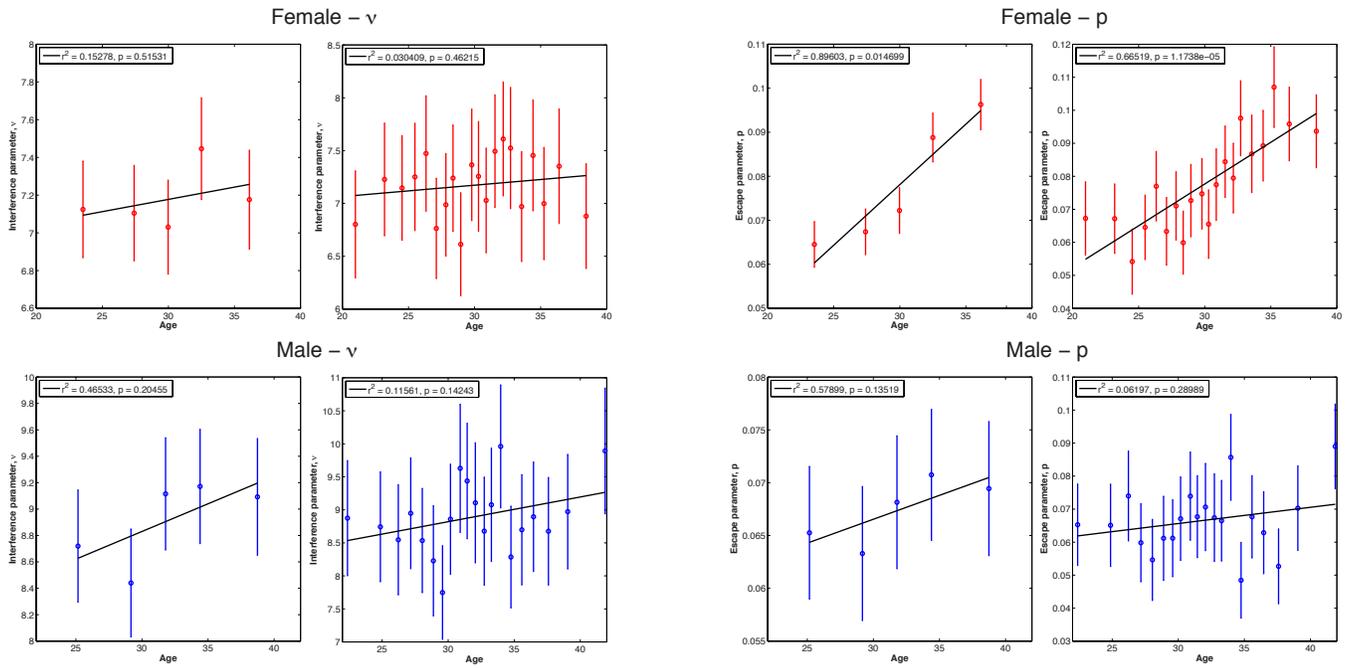

Figure S10

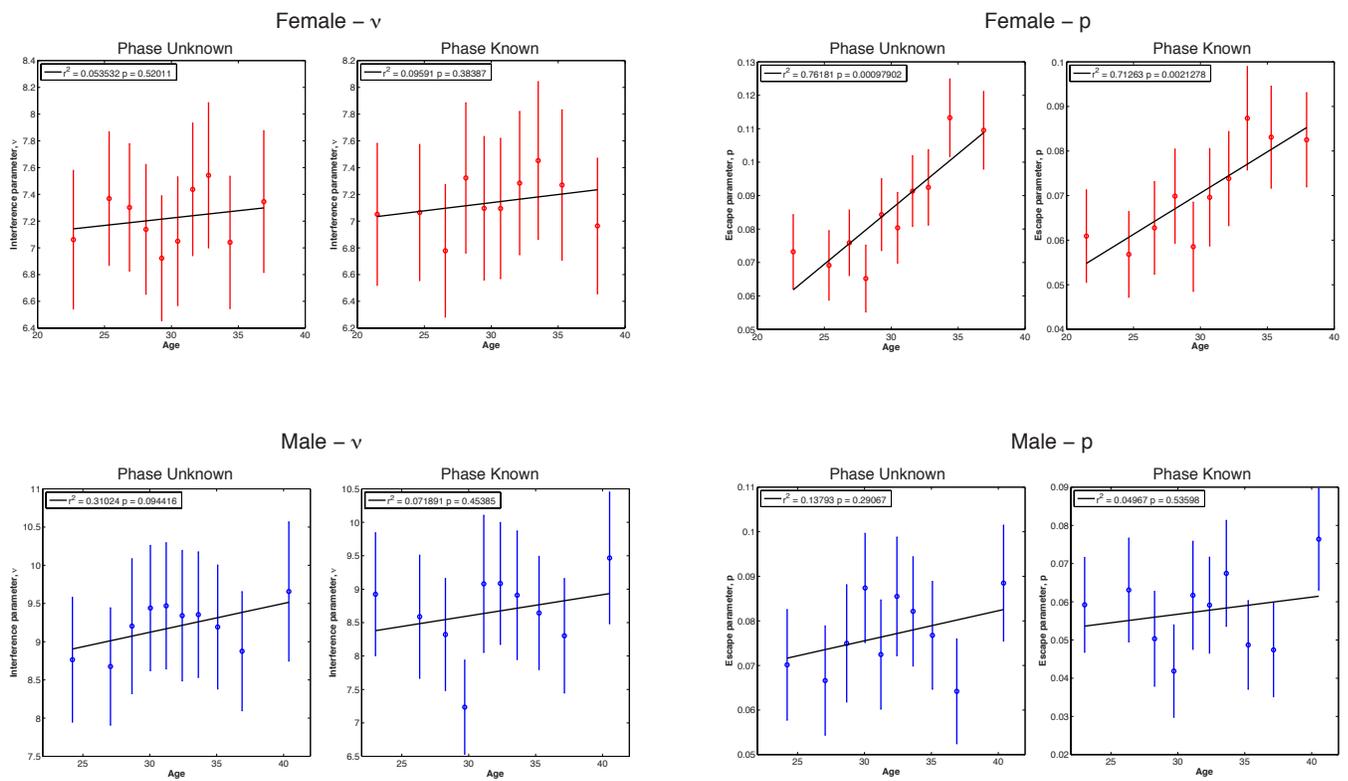

Figure S11

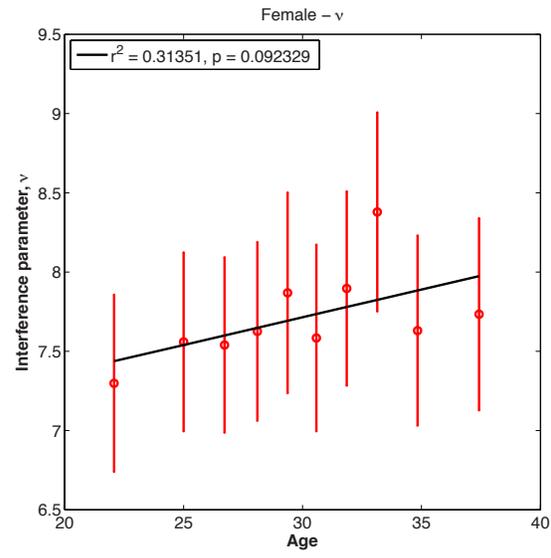
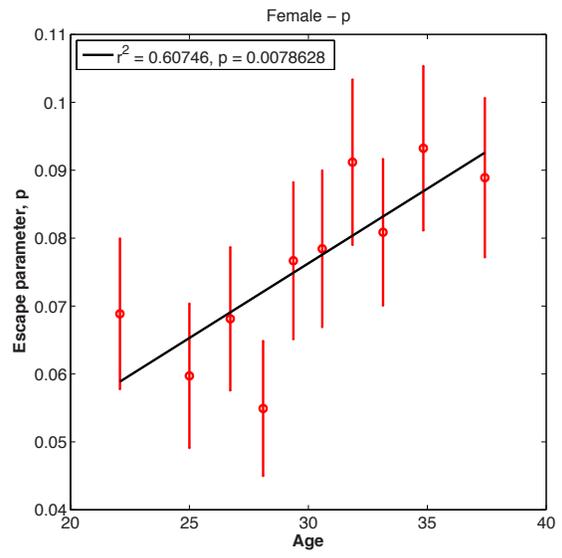
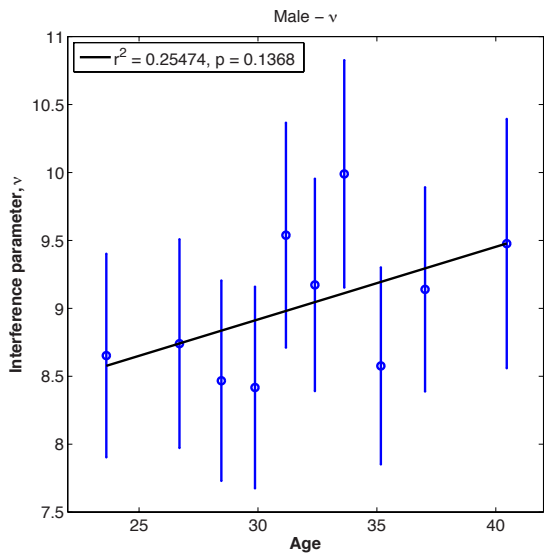
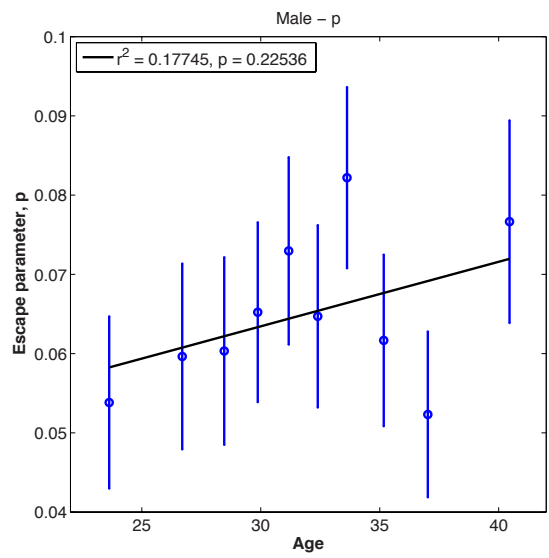

Figure S12

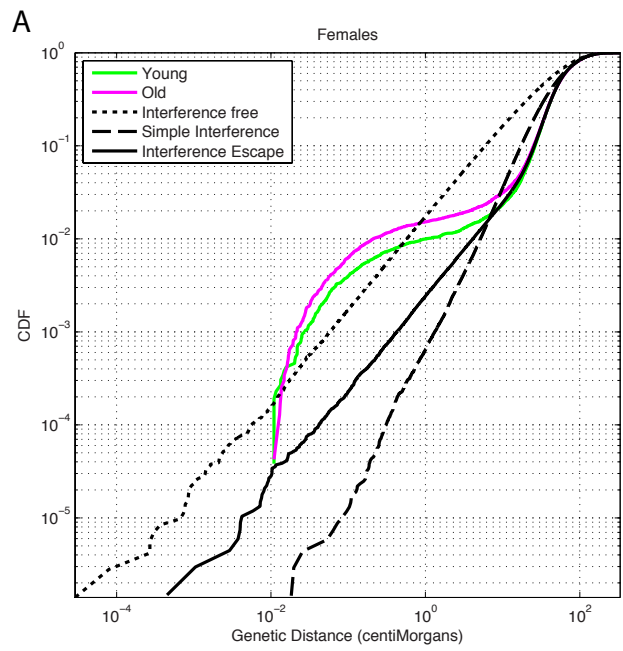 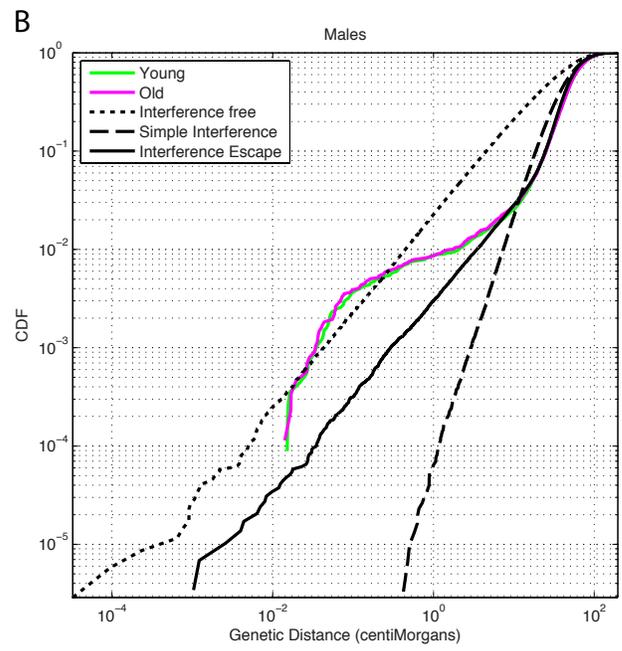

Figure S13

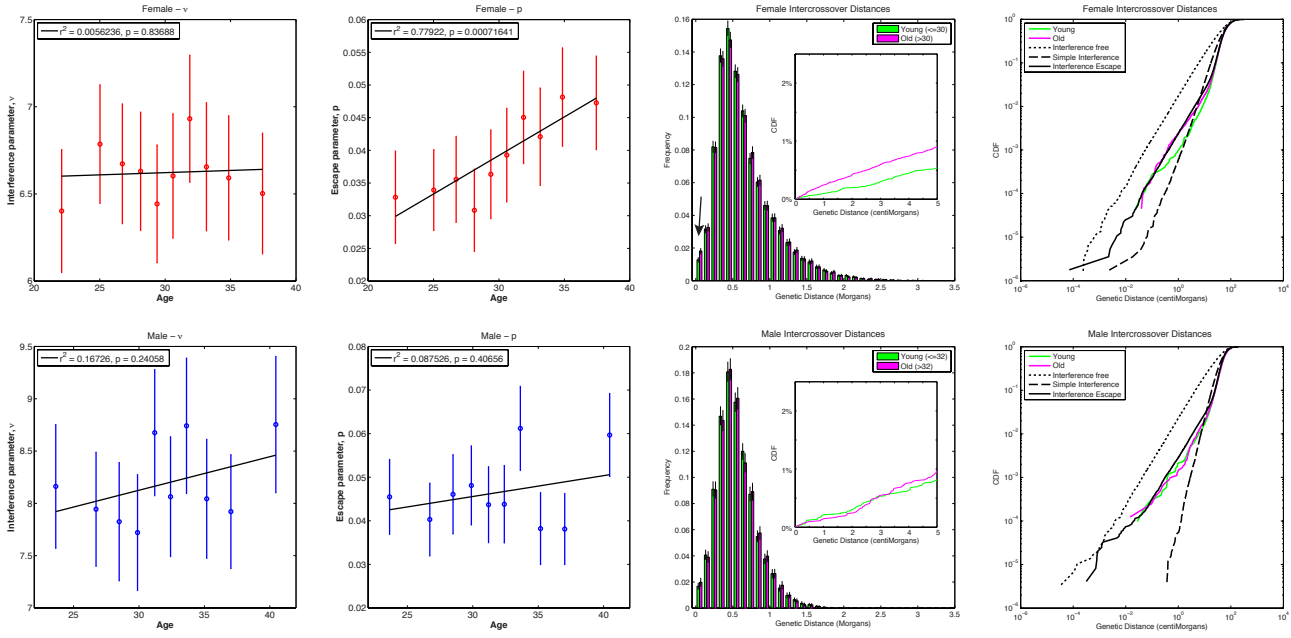

Figure S14